\documentclass[final,1p,times]{elsarticle}

\usepackage{amssymb}
\usepackage{icomma}

\usepackage{bm}
\usepackage{graphicx}
\usepackage[labelformat=simple]{subcaption}
\usepackage{xcolor}
\definecolor{LWcolor}{RGB}{0,0,255} 

\newcommand{\mycomment}[1]{}
\usepackage{icomma}

\setcitestyle{numbers,square}

\definecolor{HGcolor}{RGB}{255,20,147} 

\definecolor{LWcolor}{RGB}{0,0,255} 
\usepackage{eqnarray,amsmath}

\journal{Journal of heat and mass transfer}

\begin{document}

\begin{frontmatter}



\title{Effect of temperature-dependent thermophysical properties on turbulent forced convection under constant heat flux boundary condition}

\author[inst1]{Himani Garg}

\affiliation[inst1]{organization={Department of Energy Sciences, Lund University},
            addressline={Ole Römers väg}, 
            city={Lund},
            postcode={22100},
            country={Sweden}}
\author[inst1]{Lei Wang \corref{mycorrespondingauthor}}
\cortext[mycorrespondingauthor]{Corresponding author}
\ead{lei.wang@energy.lth.se}
\begin{abstract}
In this study, we performed highly resolved large-eddy simulations (LES) to investigate the influence of variable properties on the forced turbulent convection in a channel. The constant heat flux boundary condition permits wall temperature fluctuations and thus induces variations of fluid properties. Since the effect of viscosity on the flow exhibits $Re_\tau^{-1}$ scaling, we only considered $Re_\tau = 180$ in the present study. Compared to the flow with constant properties, results indicate that the variable properties have trivial effects on the mean velocity and temperature profiles, Reynolds shear stress, wall-normal heat flux, as well as the small-scale turbulence characteristics. However, we also observed that the turbulence intensities, low-speed streaks, burst motions, and budgets for temperature variance and wall-normal heat flux are modified by the variable properties in a perceptible way. In addition, we showed that the classic wall scaling is a good choice for flow with small and moderate variations of fluid properties.

\end{abstract}




\end{frontmatter}


\section{Introduction}
\label{sec:1}
For forced convection in a turbulent pipe or channel flow, the Nusselt number can be obtained, for example, by using the Gnielinski correlation on the basis of constant thermophysical properties \cite{Incropera2007}. However, for fluids under intensive heating, such as in the concentrating solar receivers, supercritical water-cooled reactors, heat exchangers in chemical engineering, etc., the fluid properties are temperature-dependent, which, in turn, alters the flow and temperature fields and cause a departure of heat transfer rate from the constant-property correlations. To account for the departure, Petukhov \cite{Petukhov1970} presented a theoretical analysis of heat transfer and friction to fluids with variable thermophysical properties. In this work, the Nusselt number and friction factor, which are calculated by the constant-property equations, are modified by the viscosity ratios evaluated at bulk and wall temperatures, or $Nu\/⁄Nu_{\text{const}}=(\mu_{\text{bulk}}/ \mu_{\text{wall}} )^n$ and $f/f_{\text{const}}=(\mu_{\text{bulk}}/\mu_{\text{wall}} )^{-m}$, where $n$ and $m$ depend on the specific flow conditions. Since then, further experimental and theoretical investigations have been carried out to propose different correlations which take the variable properties into account \cite{Gnielinski1976,Buyukalaca1998,Gnielinski2013,Zhao2018}. These works, however, mainly focused on the determination of coefficients $n$ and $m$, rather than on the characterization of the flow behavior from a physical point of view.

Among analytical and numerical studies dealing with variable properties problems, Pinarbasi et al. \cite{Pinarbasi2005} studied the influence of temperature-dependent viscosity and thermal conductivity on the velocity and temperature fields in a two-dimensional channel flow with isothermal boundary conditions. It was found that variable properties have some impacts on the temperature field but very scarce effects on the velocity profiles. Zonta et al. \cite{Zonta2012} performed a DNS study of the forced convection of water in a heated channel in which the viscosity of the fluid is temperature-dependent. Compared with the isothermal flow with constant viscosity, they observed that turbulence is promoted on the cold side of the channel where viscosity is higher; while turbulence is dampened on the hot side where viscosity is lower. In addition, they identified the connection between the variable viscosity and the near-wall turbulence structures that sustain the regeneration cycle, i.e., low-speed streaks, bursting events, and quasi-streamwise vortices. Similarly, using water as a working fluid, Lee et al. \cite{Lee2013} conducted DNS studies of turbulent boundary layers over isothermally heated walls in which the effect of viscosity stratification on turbulence and skin friction was studied. They demonstrated that the turbulence intensity and the Reynolds shear stress were decreased in the heated walls. In line with the attenuated turbulence transport, the skin-friction factor was reduced by up to 26\%. Subsequently, the same author, Lee et al. \cite{Lee2014}, found that the Stanton number and the Reynolds analogy factor are augmented by 10\% and 44\%, respectively, in the variable viscosity flow due to the modification of the near-wall turbulence. In strongly anisothermal wall-bounded air flows, the momentum and energy exchange are also under the influence of variations in temperature. Contrary to the high-speed compressible flow, acoustic waves have a negligible impact in anisothermal flows with low Mach numbers. Serra et al. \cite{Serra2012} performed thermal large-eddy simulations (LES) in a channel flow asymmetrically heated from both sides. They showed that the impact of temperature-dependent viscosity and thermal conductivity is less important than the impact of variable density which is caused by thermal expansion. Toutant and Bataille \cite{Toutant2013} implemented DNS studies of forced convection in an anisothermal channel flow with highly variable fluid properties. Their data illustrated that the turbulence intensity and Reynolds shear stress are enhanced on the cold side; while on the hot side, the increase of viscosity and thermal expansion are in competition to affect the turbulent motions. To assess the isolated influence produced by the temperature-dependent viscosity and thermal conductivity, Wang et al. \cite{Wang2019} performed LES to study the turbulent heat transfer in an anisothermal channel flow. Their data showed that variation in viscosity and thermal conductivity leads to the modification of the mean velocity and temperature fields. However, the molecular momentum and heat transport across the channel remain unaffected. Consistent with the results in \cite{Toutant2013}, they also observed that the turbulence is promoted at the cold side even though variable properties have negligible effect in the near-wall region $y^+ < 20$.

More recently, there has been an increased interest in studying turbulent heat transfer to fluids at supercritical pressures \cite{Yoo2013,Peeters2016,Nemati2016}. These fluids exhibit strong thermophysical property variations as the pseudo-critical temperature is crossed. Yoo \cite{Yoo2013} pointed out that the two dominant mechanisms in modulating turbulence and scalar transport in supercritical fluids are buoyancy and large variations of properties. Peeters et al. \cite{Peeters2016} conducted DNS of supercritical fluid in an annular channel with a hot inner wall and a cold outer wall. It was shown that the turbulence was attenuated at the hot wall but amplified at the cold wall. It is worth noting that this trend is similar to the sub-critical fluid in \cite{Zonta2012}. However, by analyzing a transport equation for the coherent streak flank strength, it was found that thermophysical property fluctuations significantly affect streak evolution and the self-regeneration process in near-wall turbulence. Nemati et al. \cite{Nemati2016} studied the effect of thermal boundary conditions on turbulent heat transfer to fluids at supercritical pressures. Two different thermal wall boundary conditions were considered: the first corresponds to a Neumann boundary condition that permits wall temperature fluctuations; the second corresponds to a Dirichlet boundary condition that does not allow wall temperature fluctuations. The results showed an enhancement of 7\% in Nusselt number under the Neumann boundary condition because large property variations are induced that consequently cause an increase of wall-normal velocity fluctuations and, thus, an increased overall heat flux. 

In this work, we aim to investigate the influence of temperature-dependent viscosity and thermal conductivity on turbulent forced convection under the constant heat flux boundary condition (or Neumann boundary condition). To this end, we performed highly-resolved LES studies in a heated channel flow where the air is treated as an incompressible fluid and  isolated from gravity. The variable viscosity and thermal diffusivity are connected by the Prandtl number, which is kept constant at 0.71 in our study. To highlight the viscous effect, we focused on $Re_\tau=180$ because the viscous term exhibits a $Re_\tau^{-1}$ scaling in our methodological framework. Under the constant heat flux boundary condition,  enthalpy (or, the bulk-mean temperature) usually undergoes a linear increase from the inlet to the exit. This makes it hard to implement the periodic boundary condition in the computational domain. Therefore, a modification of temperature is introduced into the energy equation to accommodate this change. This part is described in Section 2. Results and discussions are presented in Section 3. Finally, the conclusions are given in Section 4.
\section{Numerical methodology}
\label{sec:2}
\begin{figure}[htb!]%
      \centering
      \includegraphics[width=0.8\linewidth]{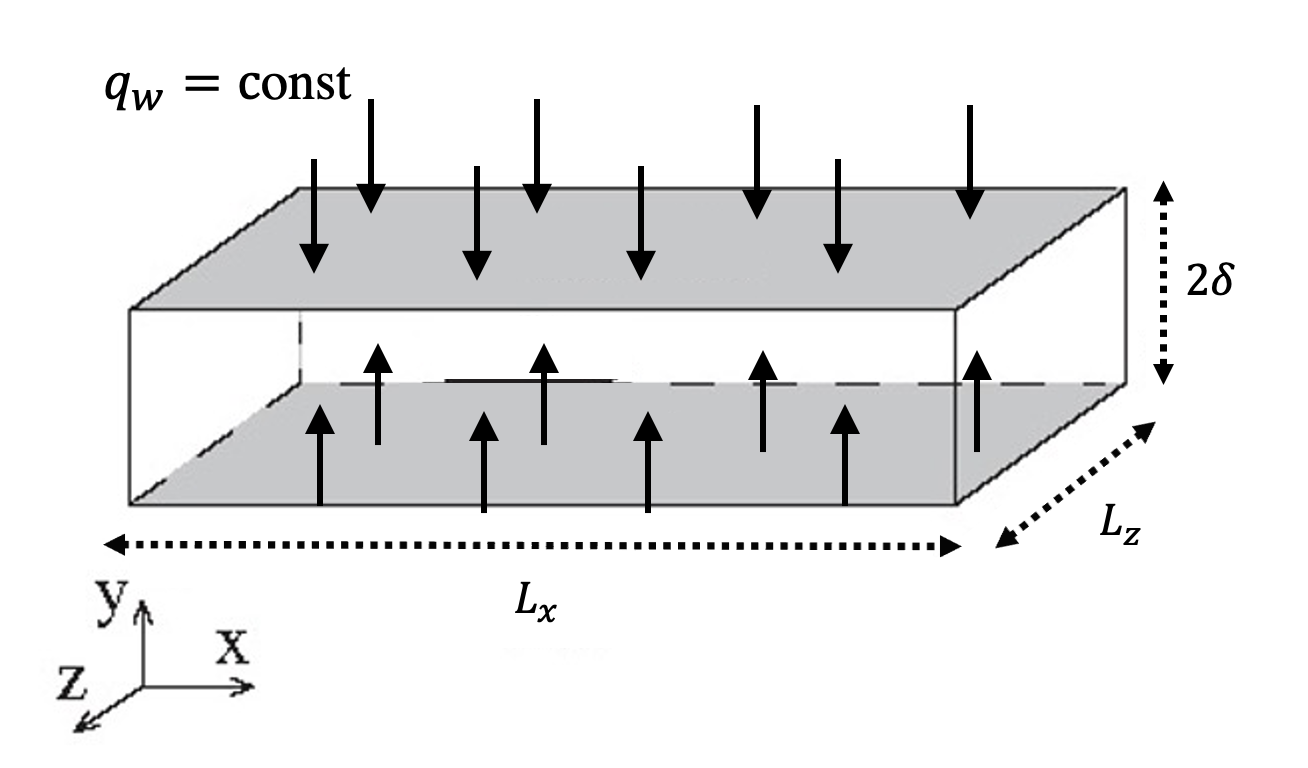}
   \caption{Schematic of channel flow configuration.}
   \label{fig: 1}
\end{figure}

In this study,  wall-resolved LES of thermal flow in a Poiseuille turbulent channel has been conducted within a computational box of $L_x = 2\pi\delta$, $L_y = 2\delta$, and $L_z=\pi\delta$, with spanwise and streamwise periodicity, shown in Fig. \ref{fig: 1}. The flow is considered fully incompressible, Newtonian, and heated with a uniform wall heat flux $q_w$.  
 The streamwise, wall-normal, and spanwise coordinates are $x$, $y$, and $z$, respectively and the corresponding velocity components are $u$, $v$, and $w$. No-slip boundary condition is imposed on the walls. The temperature is denoted by $T$.  Statistically, time-averaged quantities are denoted by angular brackets, $\left<...\right>$, whereas fluctuating quantities are denoted by prime, $\left(...\right)'$. In LES, the governing equations for the conservation of mass, momentum, and energy are averaged by a spatial filtering procedure, which is denoted by the symbol $\tilde{}$. After introducing the subgrid-scale (SGS) model, the filtered LES equations read 
 
\begin{equation}
    \frac{\partial \tilde{u}_i}{\partial x_i}=0
    \label{Eq: 1}
\end{equation}
\begin{equation}
    \frac{\partial \tilde{u}_i}{\partial t}+\frac{\partial \tilde{u}_i\tilde{u}_j}{\partial x_j} = -\frac{\partial \tilde{p}}{\partial x_i}+ \frac{\partial }{\partial x_j} \left[(\nu+\nu_{sgs})\left(\frac{\partial \tilde{u}_i}{\partial {x_j}}+\frac{\partial \tilde{u}_j}{\partial {x_i}} -\frac{2}{3}\delta_{ij}\frac{\partial \tilde{u}_k}{\partial {x_k}}\right)\right] +F\delta_{i1},
    \label{Eq: 2}
\end{equation}
\begin{equation}
\frac{\partial \tilde{T} }{\partial t} + \frac{\partial \left(\tilde{T}\tilde{u}_i\right)}{\partial x_i}=\frac{\partial}{\partial x_i}\left[ \left(\alpha+\alpha_{sgs}\right)\frac{\partial \tilde{T}}{\partial x_i}\right]
    \label{Eq: 3}
\end{equation}
where $u_i$ is the $i$th component of the velocity vector, $T$ is the temperature, $p$ is the fluctuating kinematic pressure, $F$ is the driving pressure gradient, $\nu$ and $\alpha$ are kinematic viscosity and thermal diffusivity, respectively. Here, $\alpha$ is connected with $\nu$ by Prandtl number which is kept constant at 0.71.
Under the constant heat flux boundary condition, the bulk mean temperature grows linearly with respect to the streamwise direction $x$, which makes it hard to implement periodic boundary conditions in the computational domain. In this context, we follow the suggestion of Kasagi \textit{et al.} \cite{kasagi1992} by introducing a transformed temperature $\Theta = T - T_w$ in the energy equation. The great advantage of this method is that $\Theta$ is periodic in the streamwise direction \cite{kozuka2009,lluesma2018}. Under the fully developed flow, the energy equation is changed to a new energy equation with an internal heat source which depends on the wall heat flux $q_w$ and the local streamwise velocity $u_1$, as shown in Eq. \ref{Eq: 4},
\begin{equation}
    \frac{\partial \tilde{\Theta}}{\partial t}+\frac{\partial \left(\tilde{\Theta} \tilde{u}_i\right)}{\partial x_i}=\frac{\partial}{\partial x_i}\left[ \left(\alpha+\alpha_{sgs}\right)\frac{\partial \tilde{\Theta}}{\partial x_i}\right]+\frac{q_w u_\tau}{\rho c_p \nu_w}\frac{ \tilde{u}_1}{ \left< U_1\right>}
    \label{Eq: 4}
\end{equation}
where density $\rho$, heat capacity $c_p$, viscosity at the wall $\nu_w$, are considered to be constant. The boundary condition for the transformed temperature is simply $\Theta = 0$ at $y = 0$ and $2\delta$.

To account for the temperature-dependent viscosity, we use a power relation to represent the dependency of viscosity on the transformed temperature $\Theta$, as shown in Eq. \ref{Eq: 5},  
\begin{equation}
\nu = \nu\left(\Theta_0\right)\left(\frac{\Theta+\Theta_0}{\Theta_0}\right)^n
    \label{Eq: 5}
\end{equation}
where the reference viscosity $\nu(\Theta_0 )  = 2\times10^{-5}$ m$^2$/s, and the exponent $n = 0.2$. For the case of the constant properties (CP), the viscosity is uniform throughout the domain ($\nu = 2\times10^{-5}$ m$^2$/s). For the case of the variable properties (VP), the viscosity at the wall is kept at the reference value at $2\times10^{-5}$ m$^2$/s, while the bulk-mean viscosity $\nu_b$ in the domain is about $2.4\times10^{-5}$ m$^2$/s. More detailed simulation parameters are shown in Table 1. For both considered cases, $Re_\tau = 180$. However, due to the different bulk velocity and viscosity, the bulk Reynolds numbers $Re_b$ for the CP and VP cases are 5967 and 4912, respectively. Nevertheless, they are considered to be fully turbulent flows.  

%
\begin{table}[!h]
\centering
\caption{Summary of simulation parameters.}
\label{Table 1}
\begin{tabular}{|l|l|l|l|l|l|}
\hline
Case & $U_b$ (m/s) & $u_\tau$ (m/s) & $\nu_b$ (m$^2$/s)    & $Re_b$ & $Re_\tau$ \\ \hline
CP   & 0.0597      & 0.0036         & $2.0 \times 10^{-5}$ & 5967   & 180       \\ \hline
VP   & 0.0587      & 0.0036         & $2.4 \times 10^{-5}$ & 4912   & 180       \\ \hline
\end{tabular}
\end{table}

In the simulations, the subgrid viscosity $\nu_{\text{sgs}}$ is calculated by means of the Smagorinsky model \cite{smagorinsky1963}. However, the standard Smagorinsky model results in the non-zero value of $\nu_{\text{sgs}}$ at the solid boundary. This defect can be avoided by introducing a van-Driest damping function \cite{vanDriest1956} into the length scale. The turbulent heat flux is modeled with a gradient transport model \cite{moin1991}, with $\alpha_{\text{sgs}}=\nu_{\text{sgs}}/Pr_t$, where the turbulent Prandtl number is equal to $Pr_t=0.85$.

We carried out wall-resolved LES with an open-source OpenFOAM  solver. Herein we limit ourselves to a summary of the numerical method. The governing equations are discretized on a collocated grid system. To avoid spurious pressure-velocity coupling, the Rhie-Chow interpolation is adopted \cite{rhieChow1983}. A second-order implicit scheme implements time integration. The convective terms for the momentum and energy equations are discretized with a second-order Gauss limited linear scheme \cite{sweby1984}. The Poisson equation is solved for the pressure following both the predictor and corrector steps to satisfy the continuity equation at each step. Once the velocity field is known, the temperature field is computed as a solution of the energy equation, and finally, viscosity is updated using Eq. \eqref{Eq: 5}.

The computational domain contains $102 \times 129 \times 128$ grid points. The mesh is considered non-uniform in the wall-normal direction and uniform in the streamwise and spanwise directions. The grid points in the wall-normal direction are determined by a hyperbolic tangent transformation \cite{moin1982}.
\begin{equation}
    y_k = H \left \{ 1+\frac{1}{a}\tanh \left[ -1+\frac{2k-1}{N_y}\right]\tanh ^{-1}(a)\right \}, k \in \left[1,N_y\right],
    \label{Eq: mesing}
\end{equation}
where $N_y=129$ is the number of grid points on the y-axis, and $a$ is an adjustable parameter depending on the mesh dilation. In the present case, $a=0.96135$. The dimensionless mesh sizes are $\Delta_x^+ =11.5, \Delta_z^+=4.5$, and $\Delta_y^+=0.2$ at the wall to $5.7$ at the center. These values are substantially lower than the recommendations by Piomelli \& Chasnov \cite{piomelli1996} for wall-resolved LES.


For the characterization of present simulations, classical wall scaling is used. In particular, the friction velocity, $u_\tau$, is obtained by solving the near-wall mean velocity gradient,
\begin{equation}
    u_\tau = \sqrt{\nu_w \left( \frac{\partial U}{\partial y}\right)_w},
    \label{Eq: friction velocity}
\end{equation}
where the subscript $w$ refers to the wall quantities. The friction temperature, $T_\tau$ is defined as:
\begin{equation}
    T_\tau = \frac{q_w}{\rho c_pu_\tau},
    \label{Eq: Friction temperature}
\end{equation}
where $q_w$ is the wall heat flux. Moreover, $y^+$ is defined as:
\begin{equation}
    y^+ = \frac{yu_\tau}{\nu_w}.
    \label{Eq: yPlus}
\end{equation}
In the present study, the flow remains turbulent for all the considered cases, and the turbulent Reynolds number, $Re_\tau$, is based on the $u_\tau$ and $\nu_w$:
\begin{equation}
    Re_\tau = \frac{u_\tau \delta}{\nu_w}
    \label{Eq: Friction Reynolds number}
\end{equation}

Prior to collecting data, each case has been run for approximately 100 diffusion time units ($\delta/u_\tau)$ to reach a statistically steady state which is identified by the convergence of the root-mean-square velocity ($u, v, w$) and temperature ($\theta$) fluctuations. Subsequently, data are collected over roughly 40 diffusion time units. Finally, the mean scalar quantities and various statistical moments are computed by integrating in time and in the homogeneous $x$ and $z$ directions (\textit{i.e.}, the Reynolds average). 

\section{Results and Discussions}
\label{sec:3}
\subsection{Validation}
\begin{figure}[!h]
\begin{center}
\subfloat[]{
      \includegraphics[width=0.5\linewidth]{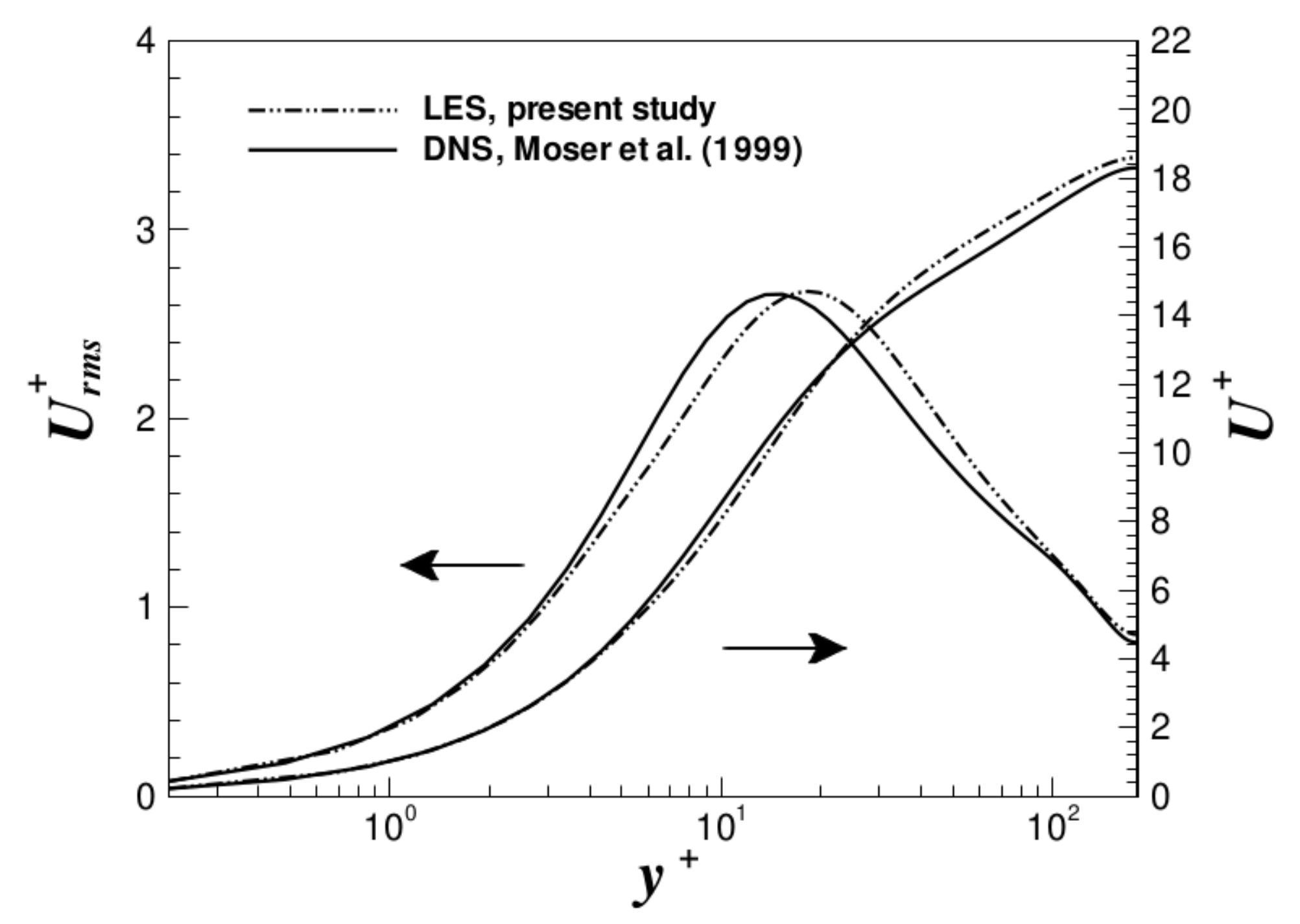}
      \label{fig: 5a}}
\subfloat[]{
      \includegraphics[width=0.5\linewidth]{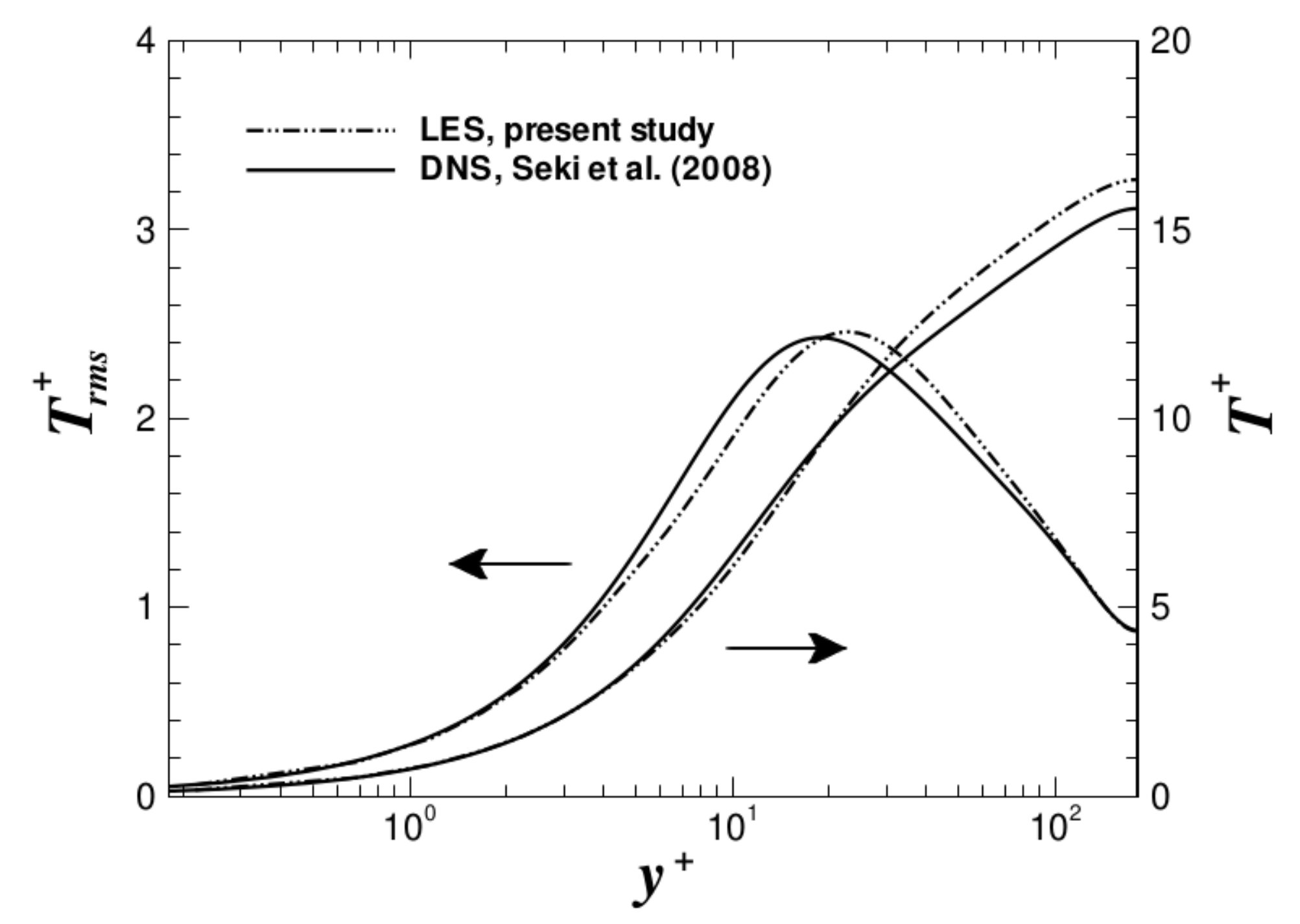}
      \label{fig: 5b}}
   \caption{ Comparison of (a) mean velocity and turbulent velocity fluctuations and (b) mean temperature and temperature fluctuations, with DNS \cite{Moser1999,Seki2003}. The results are extracted for the constant properties case.}
   \label{fig: 5}
   \end{center}
\end{figure}

Before detailing the effect of temperature-dependent viscosity on the flow properties, we performed a detailed validation against the DNS database available for the channel flow at $Re_\tau=180$ where the fluid properties are constant. Figure \ref{fig: 5a} shows the profiles for the mean streamwise velocity $U^+$  and its fluctuations $U^+_{rms}$,  plotted in semi-logarithmic coordinates. The DNS results of Moser \textit{et al.} \cite{Moser1999} are also plotted for the comparison's sake. Our simulation results in an overestimation of the mean velocity in the log region. General agreement for root-mean-square fluctuations between present simulation results and DNS is reasonably good; however, the peak value is slightly shifted outwards. In Fig. \ref{fig: 5b}, the profiles of mean temperature and its variance are also plotted. The results of Seki \textit{et al.} \cite{Seki2003} are presented for comparison. To keep consistent with \cite{Seki2003}, we use $T_{rms}^+$ instead of $\theta_{rms}^+$ to represent the temperature fluctuations. Similarly to the mean flow profiles, a slight overestimation of the results compared to DNS is observed. Like $U^+_{rms}$, the peak value of $T^+_{rms}$ is properly predicted, but the peak position is also shifted outwards. It is noted that even finer meshes would reduce the disagreement, but they cannot completely remove it. This relates to the Smagorinsky model, which reproduces the action of small scales by introducing a subgrid viscosity, but it does not model the backward of the energy cascade \cite{david2021}. Nevertheless, compared to the reference DNS data, the essential features of the flow and scalar quantities are captured, especially close to the wall, and the maximum discrepancies are within $10\%$.
\subsection{Mean flow variables}
\begin{figure}[!h]
\begin{center}
\subfloat[]{
      \includegraphics[width=0.5\linewidth]{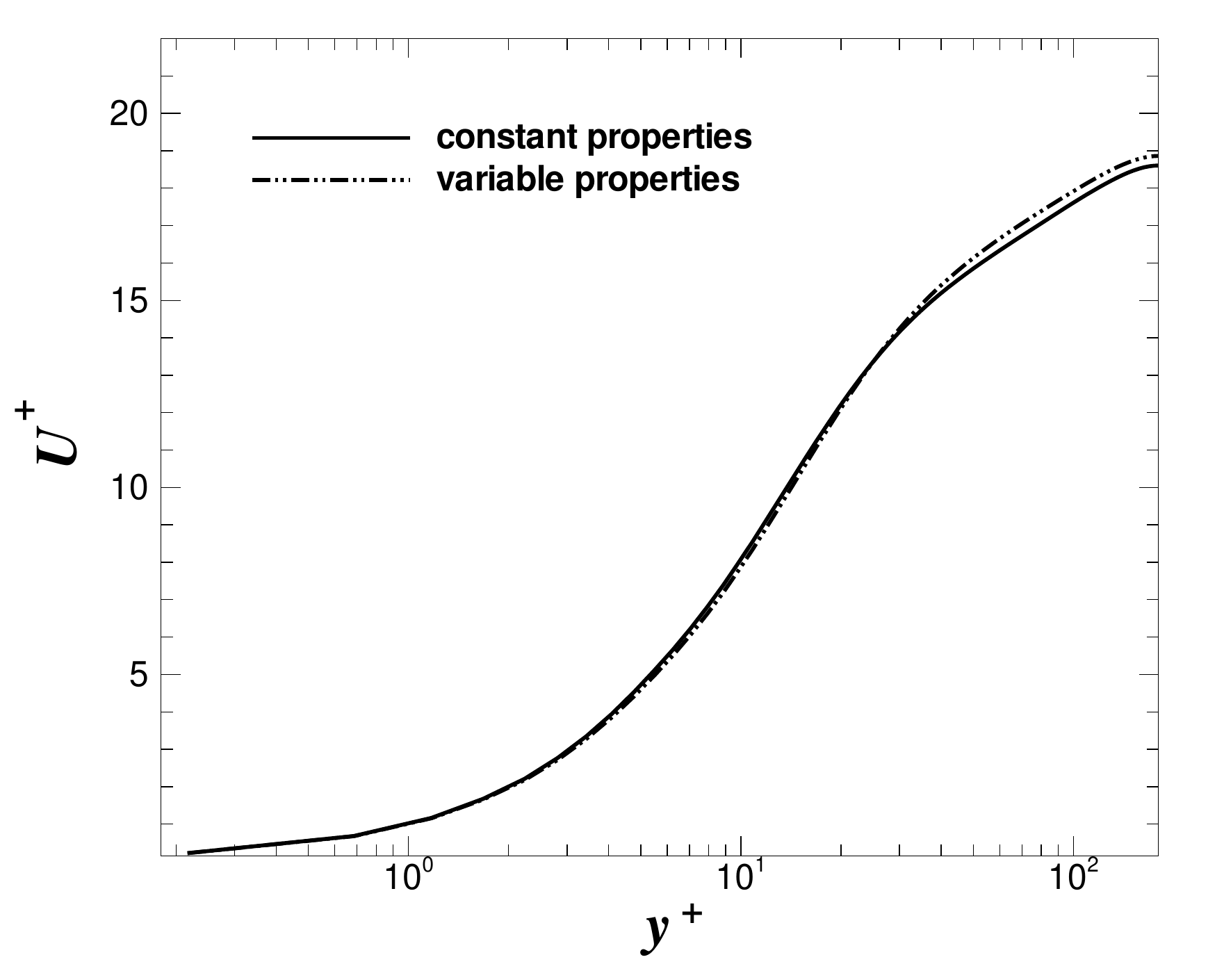}
      \label{fig: 6a}}
\subfloat[]{
      \includegraphics[width=0.5\linewidth]{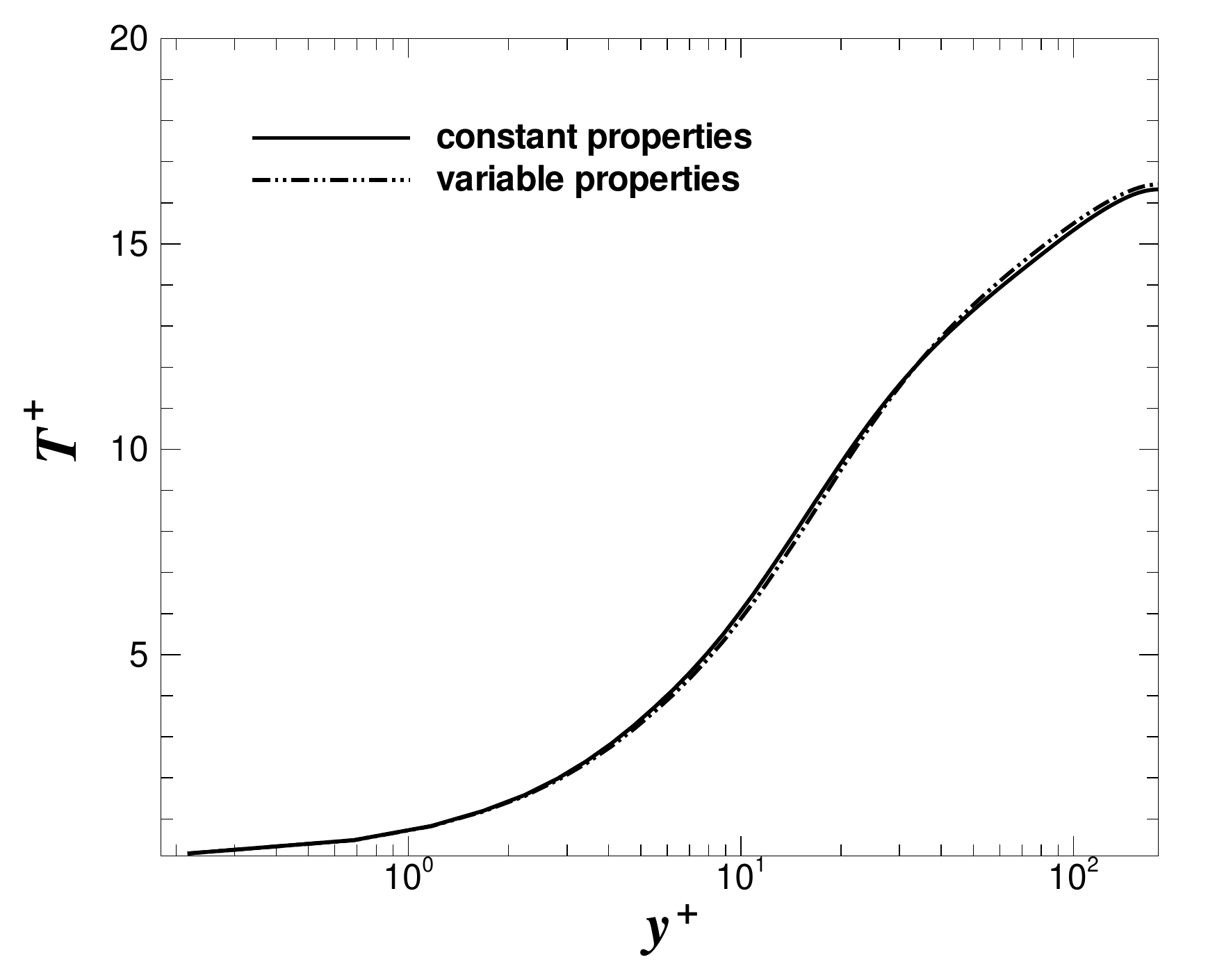}
      \label{fig: 6b}}
   \caption{ Comparison of the wall-normal distribution of (a) mean streamwise velocity and (b) temperature in the wall coordinates. The profiles of streamwise velocity and temperature are normalized by $u_\tau$ and $T_\tau$ respectively, for each case.}
   \label{fig: 6}
   \end{center}
\end{figure}

The profiles of the mean streamwise velocity and temperature plotted against the wall units are presented in Fig. \ref{fig: 6}, normalized by $u_\tau$ and $T_\tau$, respectively. In Fig. \ref{fig: 6a}, the impact of variable viscosity on the profile of the mean streamwise velocity is trivial, and only a slight difference is seen in the log region; In Fig \ref{fig: 6b}, the mean temperature profiles for the two cases agree reasonably well. This is in contrast with the results of Lee \textit{et al.} \cite{Lee2014}, showing an upward shift of mean velocity profile and a downward shift of temperature profile in the log region for the cases of  variable viscosity compared to the constant reference. The difference between \cite{Lee2014} and our studies can be attributed to the different thermal boundary conditions. In \cite{Lee2014}, the isothermal wall heating with different temperatures (or Dirichlet boundary conditions) was used. In their simulations, the higher wall temperature leads to lower viscosity of water and, thus, lower skin friction, which is associated with the upshift of the mean velocity profiles. In the present studies, the Neumann boundary condition results in the same viscosity at the wall for the two concerned cases. In particular, for the VP case, the increased viscosity in the fluid does reduce the bulk Reynolds number in the channel, as shown in Table 1, but it does not change the skin friction on the wall. In addition, we observe that the variable viscosity and thermal conductivity do not modify the mean temperature profile. This is also inconsistent with the results in \cite{Pinarbasi2005}, showing that temperature-dependent properties have some impact on the temperature field. However, in \cite{Pinarbasi2005}, they also used the isothermal boundary condition.  Despite the influence of variable properties, the good agreement of the mean velocity and temperature profiles between CP and VP cases indicates that the classic wall units are a good choice for scaling. Of course, this conclusion may only apply to the fluids having small to moderate dependence on temperature. For fluids undergoing significant variations of physical properties like superfluids, however, scaling that takes into account the temperature-dependent properties is suggested \cite{pecnik2017,patel2017}.
\subsection{Statistical properties}
\begin{figure}[!h]
\begin{center}
\subfloat[]{
      \includegraphics[width=0.5\linewidth]{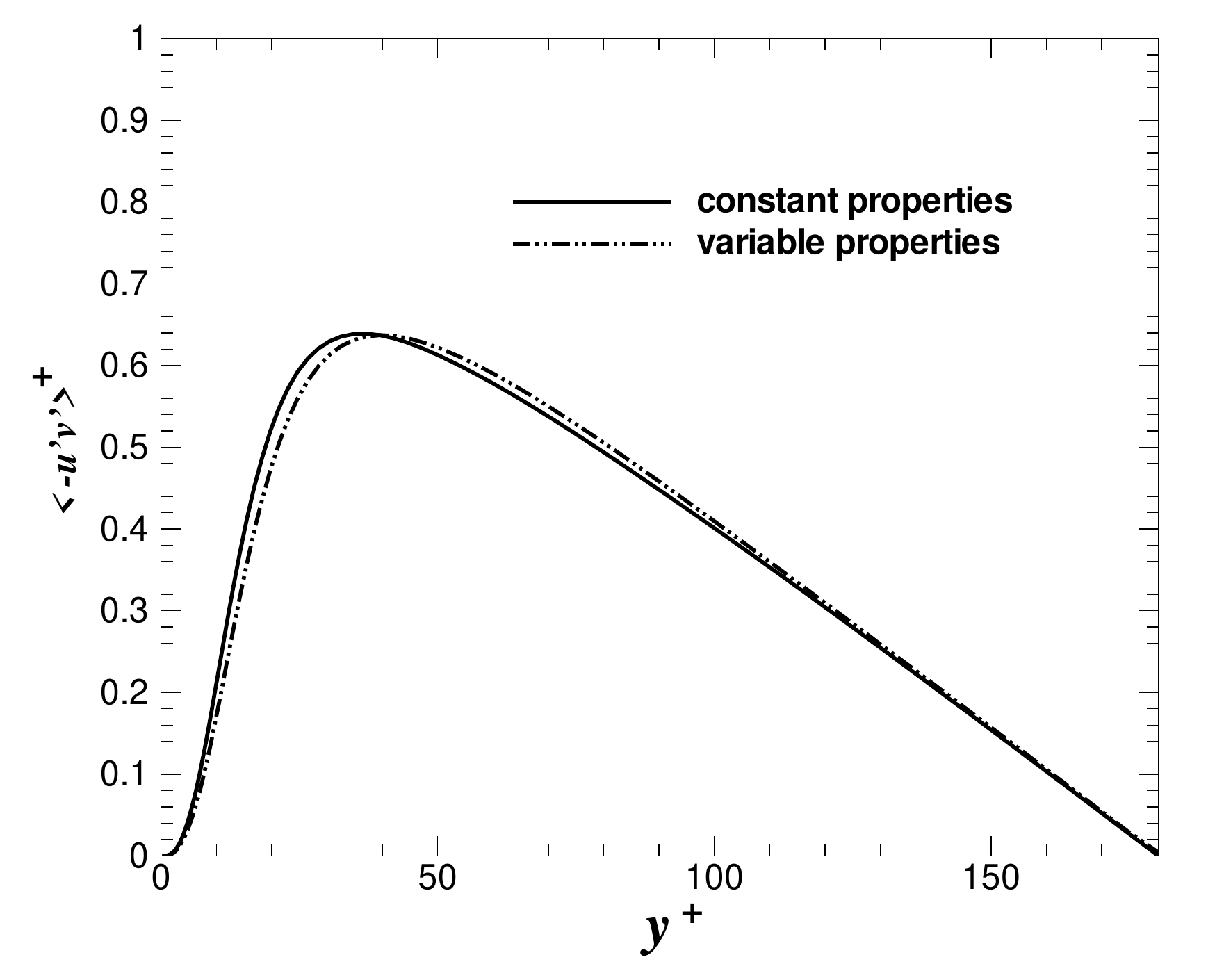}
      \label{fig: 7a}}
\subfloat[]{
      \includegraphics[width=0.5\linewidth]{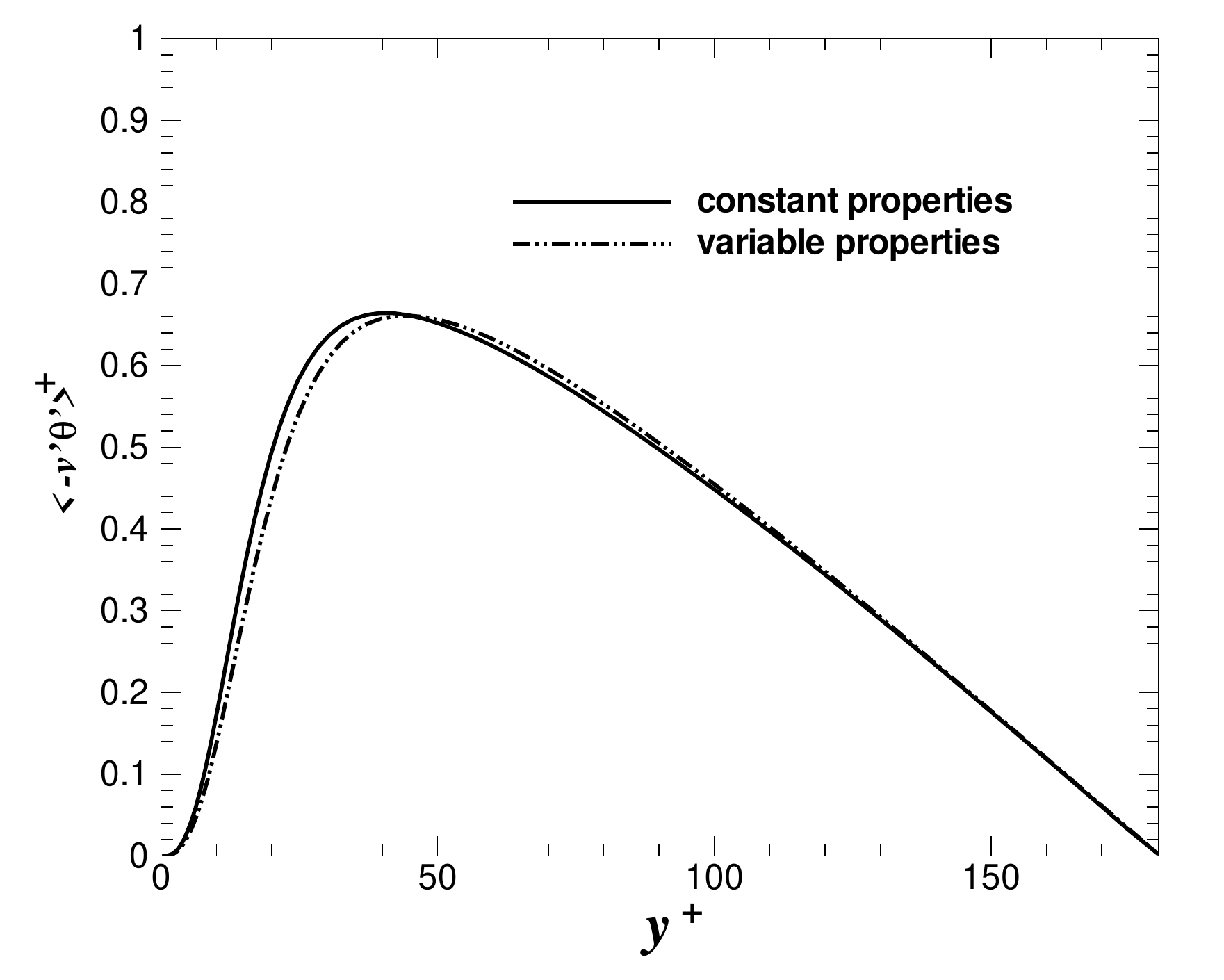}
      \label{fig: 7b}}
      
\subfloat[]{
            \includegraphics[width=0.5\linewidth]{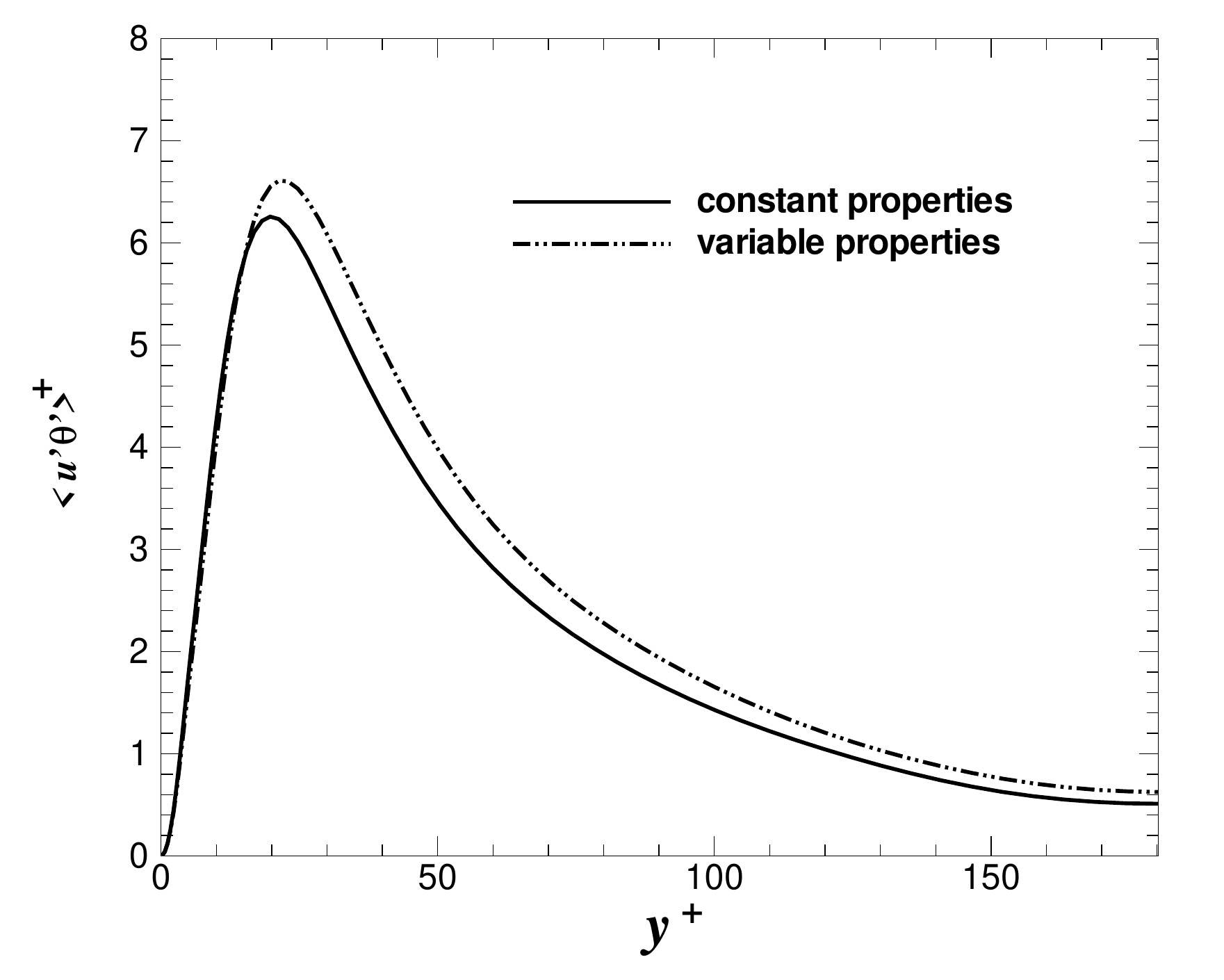}
            \label{fig: 7c}}
   \caption{ Comparison of normalized profiles of (a) turbulent shear stress, (b) wall-normal heat flux, and (c) streamwise heat flux in the wall coordinates.}
   \label{fig: 7}
   \end{center}
\end{figure}

Apart from the mean velocity and temperature profiles, it is of interest to investigate the influence of variable properties on turbulent statistical properties. In Fig. \ref{fig: 7a}, we plot the profiles of the Reynolds shear stress, $\left<-u'v' \right>$, which are normalized by $u_\tau^2$, against the wall unit $y^+$. In general, the two profiles collapse very well in the region of $y^+ >100$. Close to the wall, a small deviation of the profile with variable properties from the benchmark case is observed. In particular, the property variations slightly shift the peak position outwards but without modifying its peak value. A very similar trend is found in the profiles of wall-normal heat flux $\left<- v'\theta'\right>$, which are normalized by $u_\tau T_\tau$ and shown in Fig. \ref{fig: 7b}. The similarity between the wall-normal heat flux and the Reynolds shear stress displayed in the present study is in line with the DNS results in \cite{kong2000}, indicating a strong correlation between $u$ and $\theta$. This feature will be visited later. Fig. \ref{fig: 7c} illustrates the profiles of streamwise turbulent heat flux, $\left< u'\theta'\right>$, normalized by $u_\tau T_\tau$. It is observed that the profiles collapse very well in the wall region, $y^+ < 20$. Further away from the wall, the variable properties increase its magnitudes and shift the profile upward from the benchmark case. It is worth noting here in wall-bounded turbulent flow, the momentum and heat are mainly transported via Reynolds shear stress and wall-normal heat flux, which are almost unaffected by the property variations, as shown in the present study. Thereby, we can reach a conclusion that the variable properties have a trivial influence on the turbulent transport across the channel.

\begin{figure}[!h]
\begin{center}
\subfloat[]{
      \includegraphics[width=0.5\linewidth]{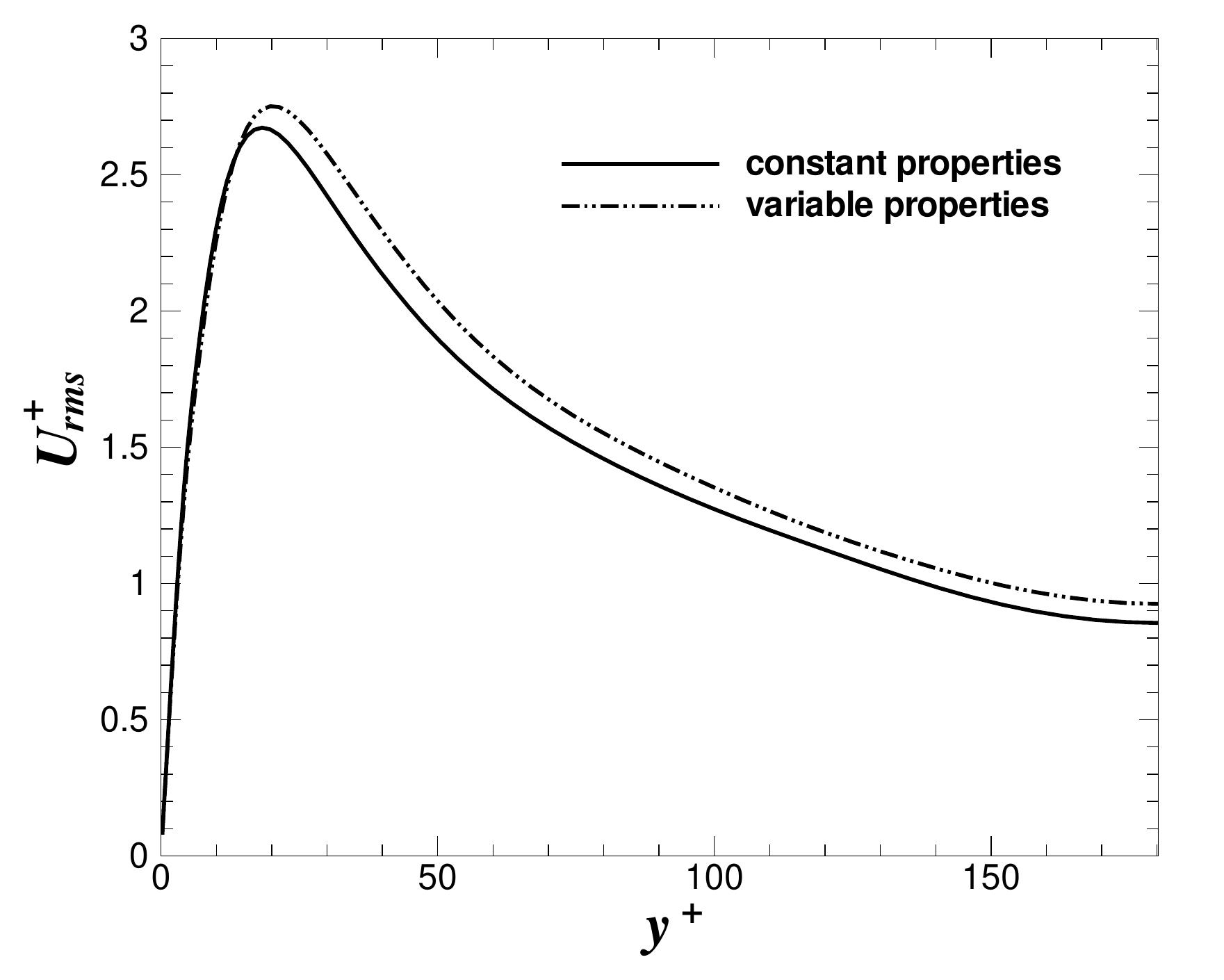}
      \label{fig: 8a}}
\subfloat[]{
      \includegraphics[width=0.5\linewidth]{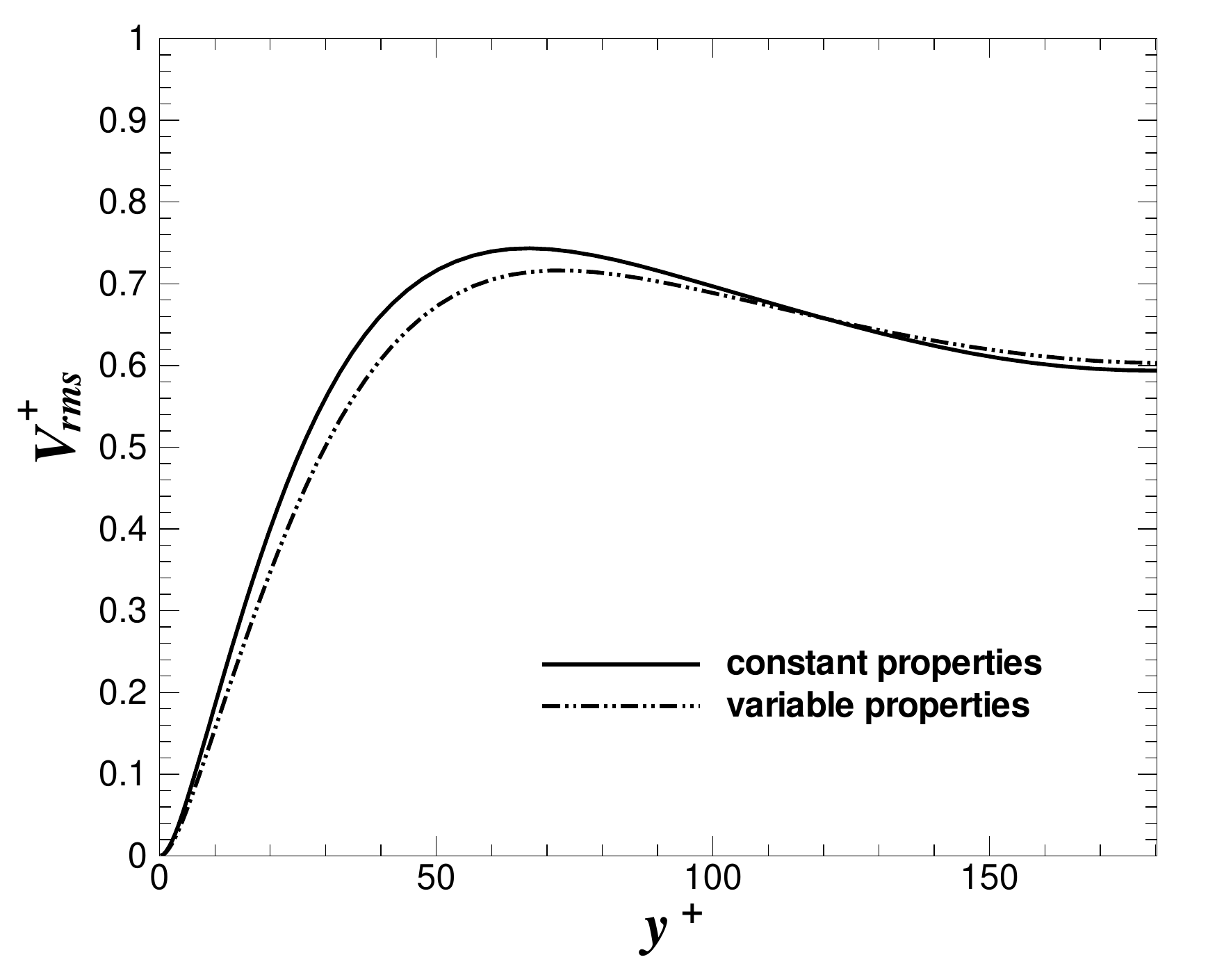}
      \label{fig: 8b}}
      
\subfloat[]{
            \includegraphics[width=0.5\linewidth]{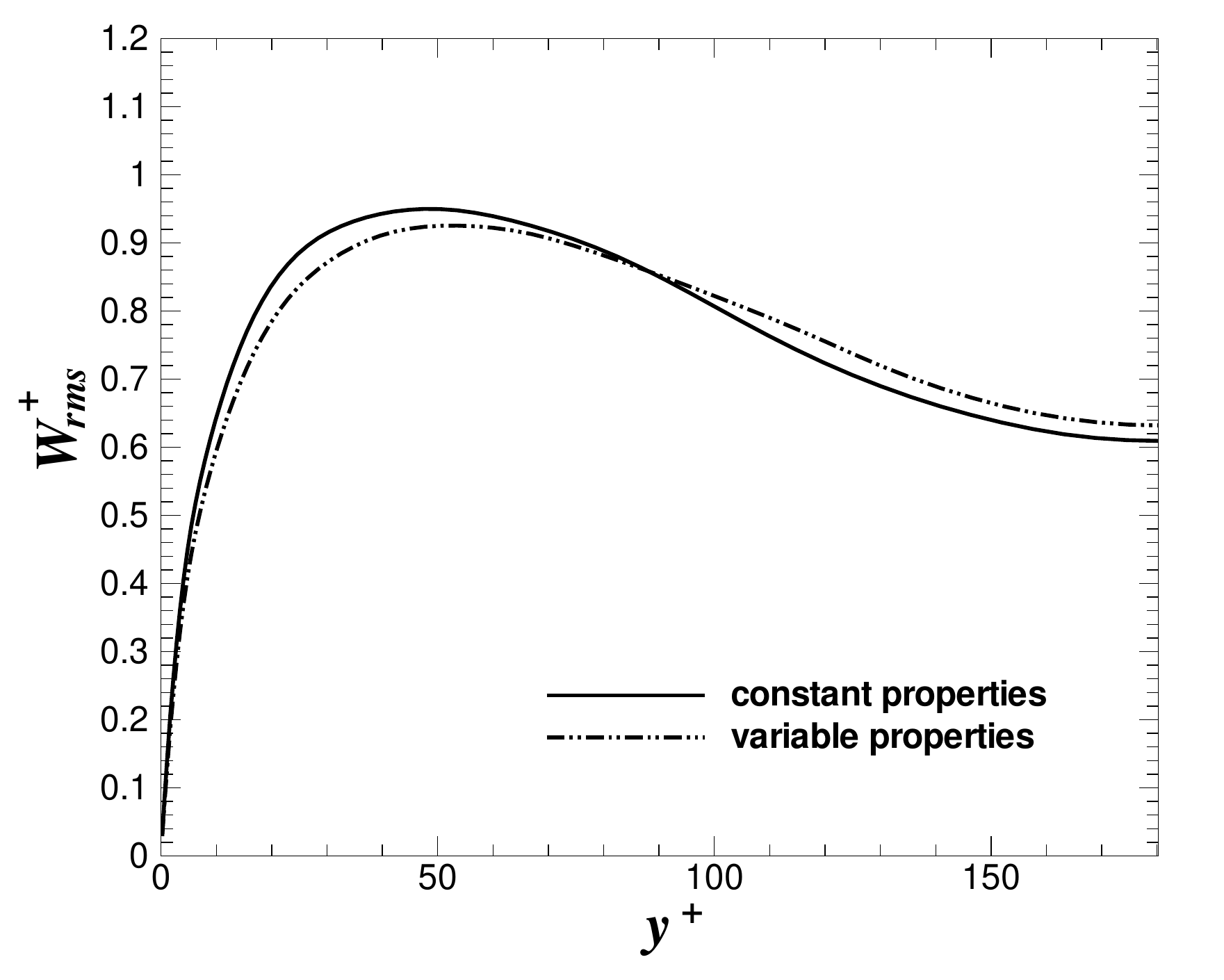}
            \label{fig: 8c}}
\subfloat[]{
            \includegraphics[width=0.5\linewidth]{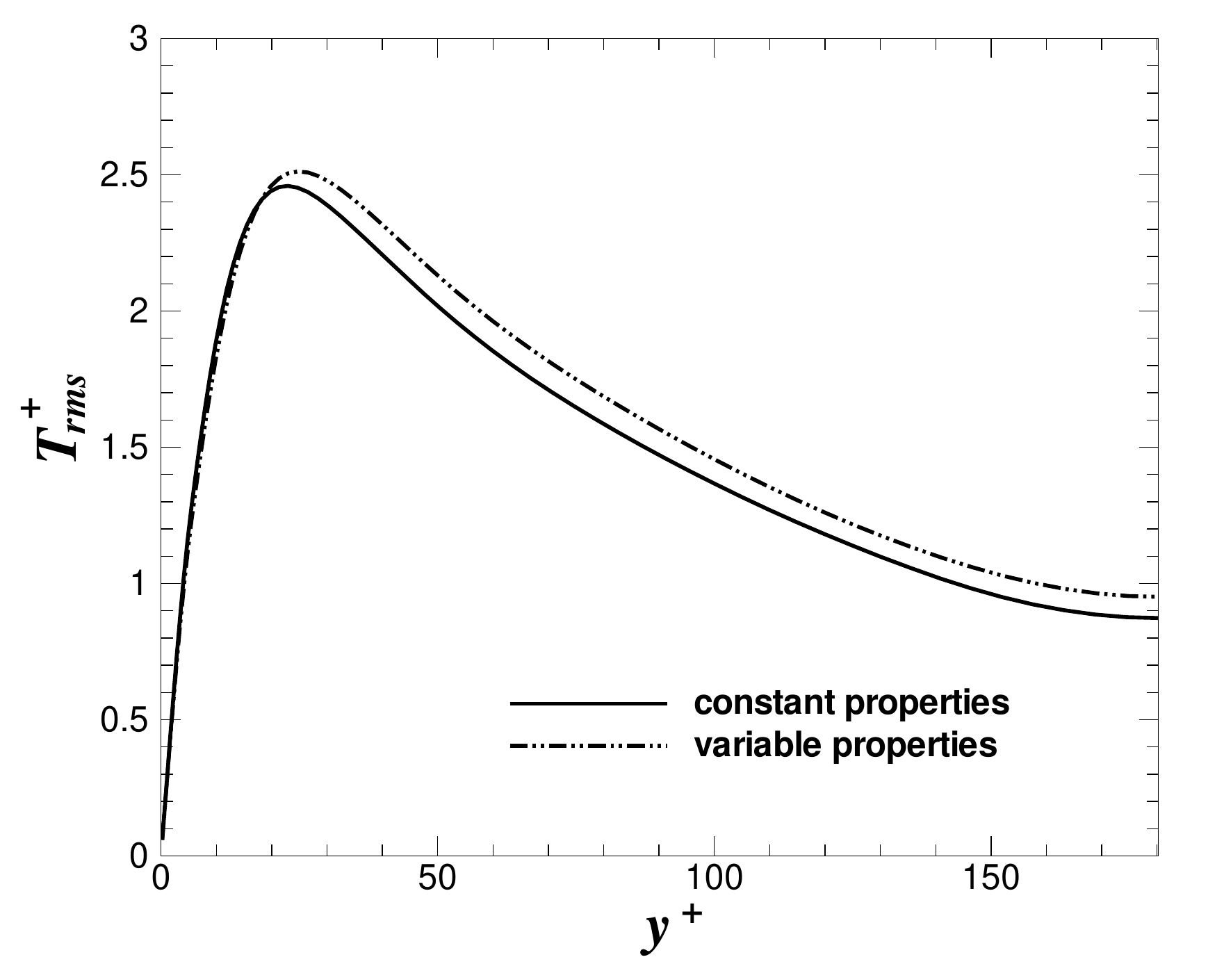}
            \label{fig: 8d}}
   \caption{ Comparison of normalized profiles of the root-mean-square of the (a)  axial velocity fluctuations, (b) wall-normal velocity, (c) spanwise velocity, and (d) root-mean-square of the temperature fluctuations, in the inner scale units.}
   \label{fig: 8}
   \end{center}
\end{figure}

To further inspect the impact of fluid properties on the turbulence intensities, the distributions of root-mean-square streamwise, wall-normal, and spanwise velocity, as well as temperature fluctuations, are presented in Fig. \ref{fig: 8}. Here, we observe the striking similarity between the streamwise velocity fluctuations $u^+_{rms}$ and temperature fluctuations $\theta^+_{rms}$. In the near-wall region $y^+ < 20$, the variable properties have no effect on the streamwise velocity and temperature fluctuations, and the profiles collapse very well for the CP and VP cases. In the outer region $y^+ > 20$, the increased viscosity and thermal conductivity tend to enhance the magnitudes of $u^+_{rms}$ and $\theta^+_{rms}$, respectively. This is also in contrast with the results of Lee \textit{et al.} \cite{Lee2014}. In \cite{Lee2014}, the wall heating results in the diminished turbulence intensity; however, if normalized by the friction velocity $u_\tau$, a better agreement was achieved for various  $u^+_{rms}$ profiles. Further inspection of Fig. \ref{fig: 8} shows that the variable properties have some visible impacts on the quantities of $v^+_{rms}$ and $w^+_{rms}$. In particular, close to the wall region, both $v_{rms}$ and $w_{rms}$ levels are suppressed compared to the benchmark case, and further away from the wall, the fluctuation level enhances. In other words, $v^+_{rms}$ and $w^+_{rms}$ can feel the influence of the variable properties throughout the channel, while $u^+_{rms}$ and $\theta^+_{rms}$ are modified only in the outer region $y^+ > 20$. 
\subsection{Probability density functions}
\begin{figure}[!h]
\begin{center}
\subfloat[]{
      \includegraphics[width=0.5\linewidth]{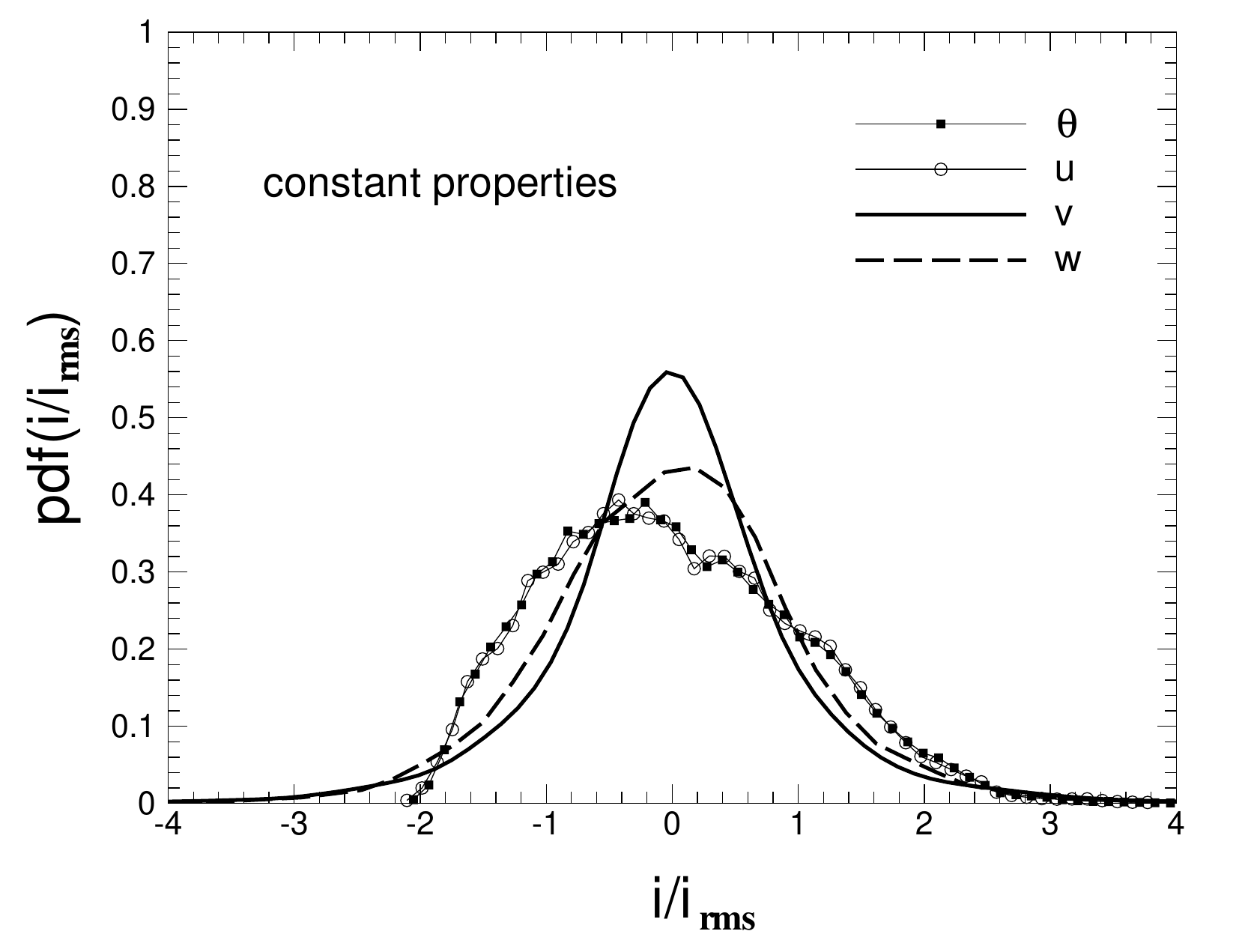}
      \label{fig: 9a}}
\subfloat[]{
      \includegraphics[width=0.5\linewidth]{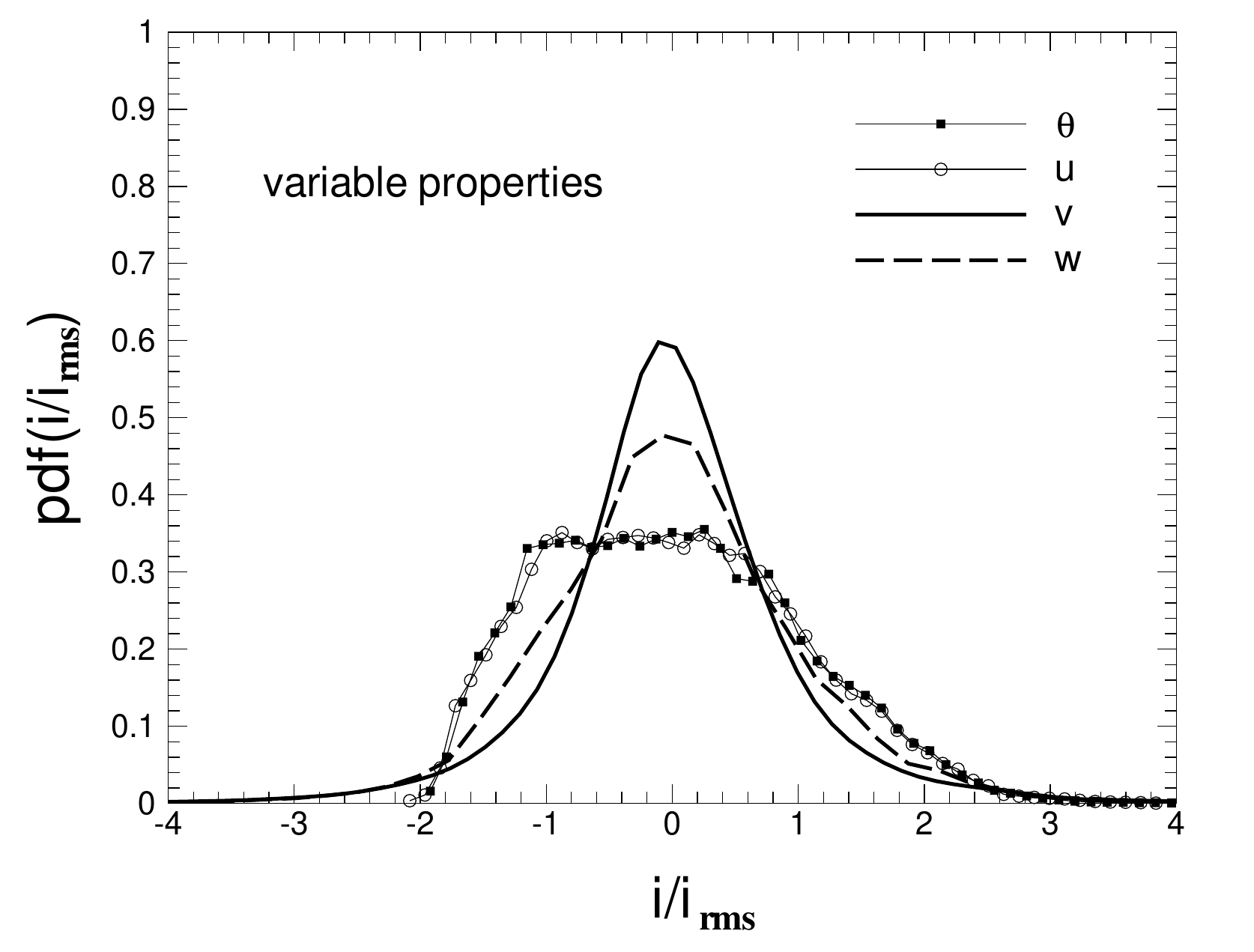}
      \label{fig: 9b}}
   \caption{Probability density function at $y^+ = 5.5$.}
   \label{fig: 9}
 \end{center}  
\end{figure}

Probability density functions (PDFs) of velocity and scalar quantities such as temperature are fundamental in studying and modeling statistical characteristics of flow and heat transfer in wall-bounded turbulent flows. The PDFs of streamwise, wall-normal, and spanwise velocity components and of temperature $\theta$ are depicted in Fig. \ref{fig: 9} for both constant properties (see Fig. \ref{fig: 9a}) and variable properties cases \ref{fig: 9b}. To highlight the dynamic structures in the near-wall region, the PDF profiles are plotted only at the position  $y^+=5.5$. By inspecting Fig. \ref{fig: 9a}, the first impression conveyed to us is that PDF profiles of $u$ and $\theta$ are in good agreement, indicating that the two quantities are strongly coupled; Second, the negatively skewed peak of $u$ profile suggests that low-speed streaks are dominant structures in the near-wall sublayer. From a heat transfer point of view, the low-speed streaks mainly contain the colder fluid. In addition, both PDF profiles for $u$ and $\theta$ show a long tail in the positive region, indicating that the flow in the vicinity of the wall is of intermittent nature. These predictions are in accordance with the experimental results of Antonia \textit{et al.} \cite{antonia1988} and the DNS results of Redjem-Saad  \textit{et al.} \cite{redjem2007}. Under the variable properties, as shown in Fig. \ref{fig: 9b}, the strong coupling between $u$ and $\theta$ is not affected, however, the negatively skewed peak is replaced by a plateau which keeps being negatively inclined. Regarding the other two velocity components, under constant properties, the PDF profile of $w$ displays a quasi-Gaussian distribution, while $v$ profile exhibits a higher peak than $w$. This is due to the fact that approaching towards the wall, $v$ is proportional to $y^3$ while $w$  ${\displaystyle \propto }$ $y^2$ \cite{pope2000}. Under the variable properties, as shown in Fig. \ref{fig: 9b},  the PDF profiles still keep their symmetrical shape; however, the peak value is seen to increase, especially for the $w$ component. This indicates that the distribution of the spanwise velocity loses its Gaussian nature under the influence of variable properties. 
\subsection{Joint probability distribution function}
 \begin{figure}[]
\begin{center}
\subfloat[]{
      \includegraphics[width=0.48\linewidth]{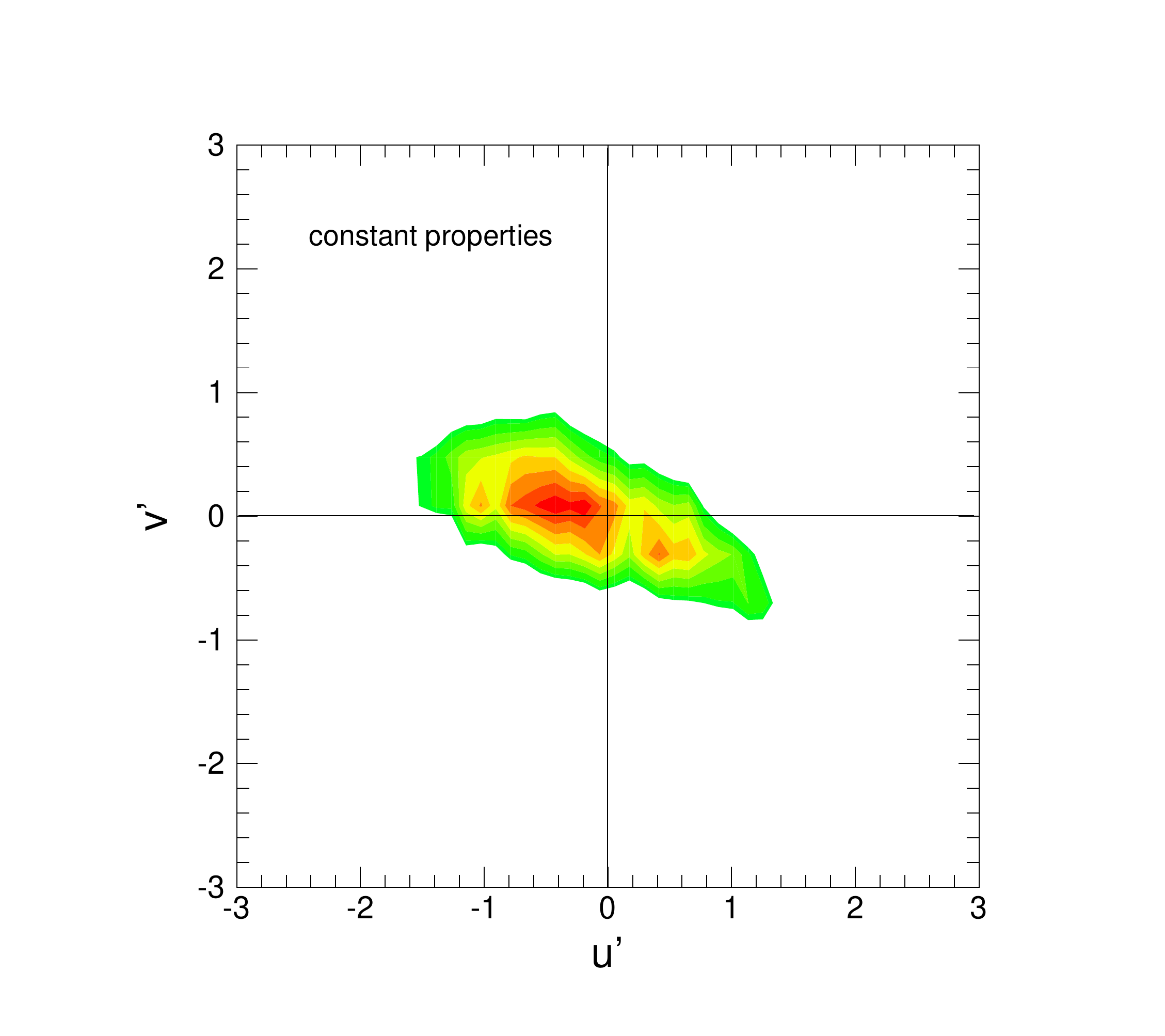}
      \label{fig: 12a}}
\subfloat[]{
      \includegraphics[width=0.48\linewidth]{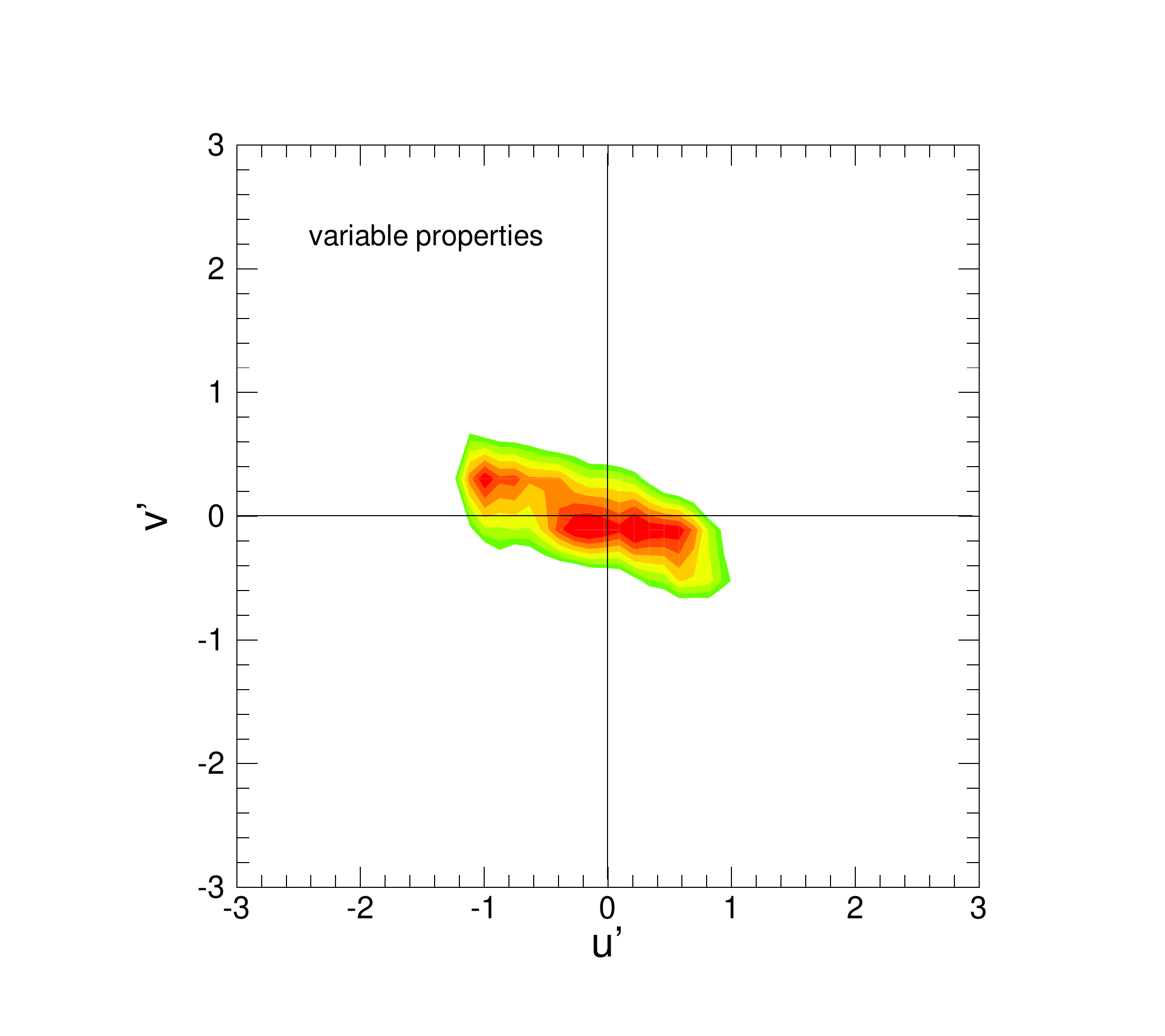}
      \label{fig: 12b}}
    
\subfloat[]{
\hspace{1.2em}        \includegraphics[width=0.48\linewidth]{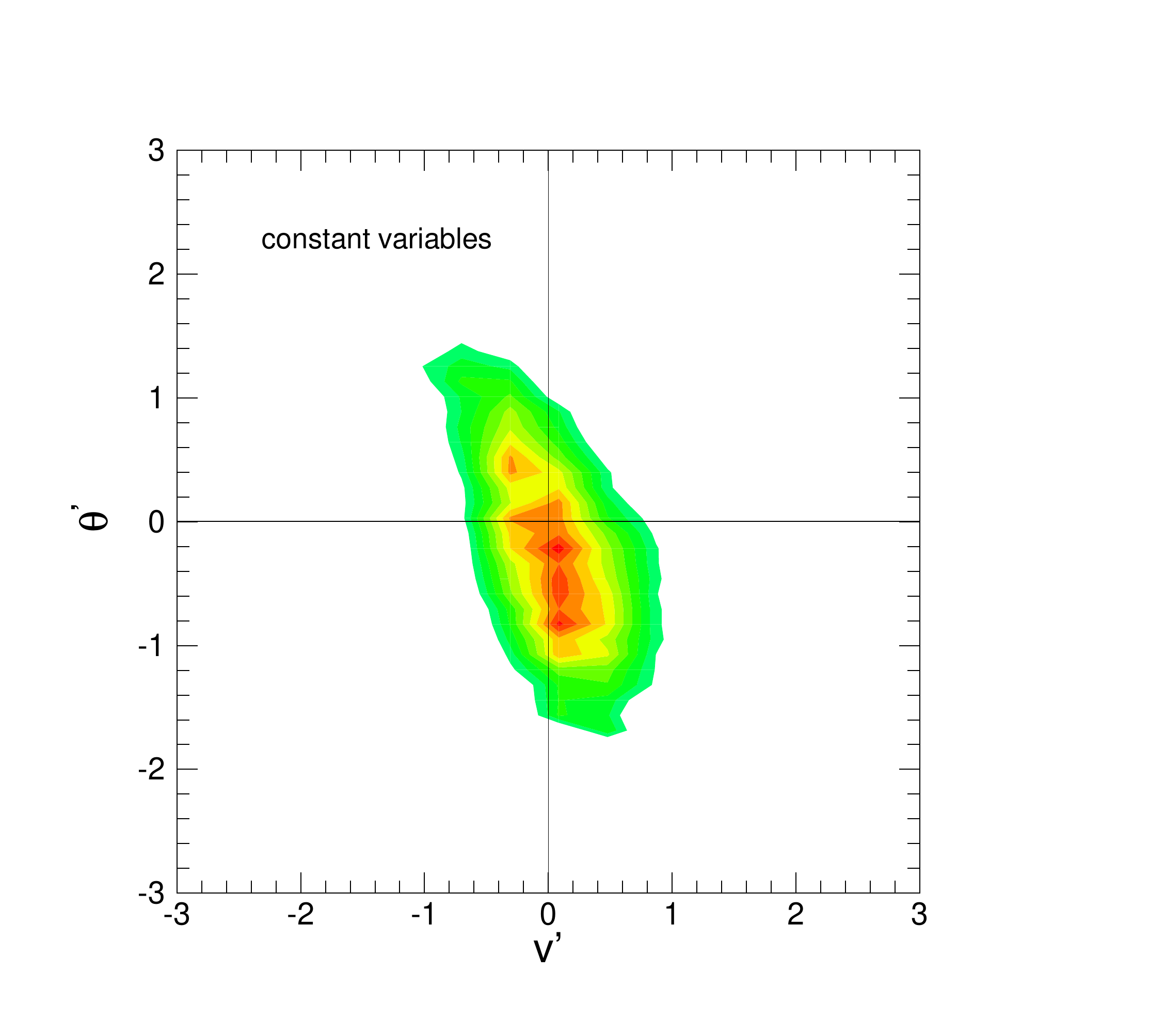}
      \label{fig: 12c}}     
\subfloat[]{
      \includegraphics[width=0.48\linewidth]{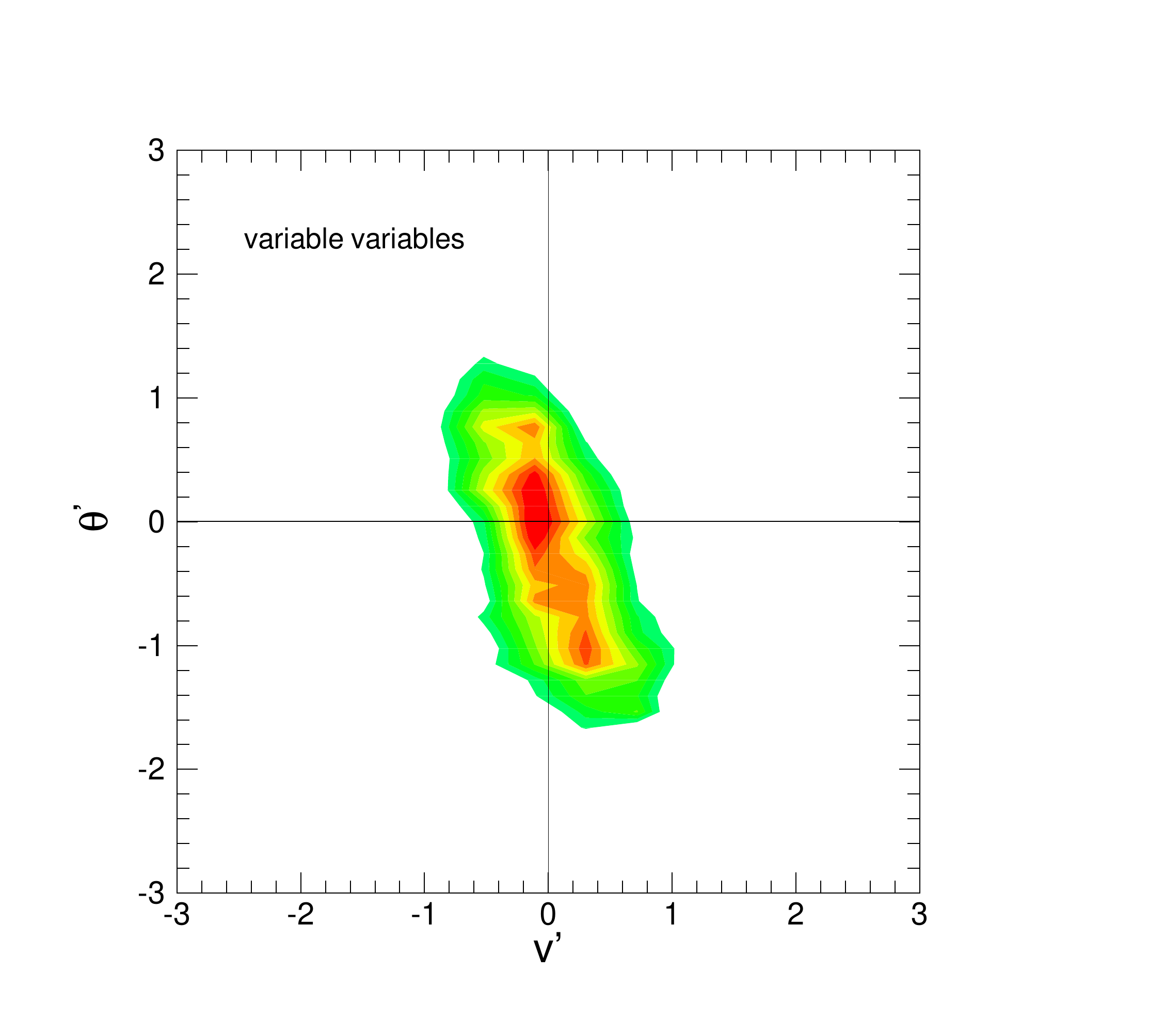}
      \label{fig: 12d}}
      
\subfloat[]{
      \includegraphics[width=0.48\linewidth]{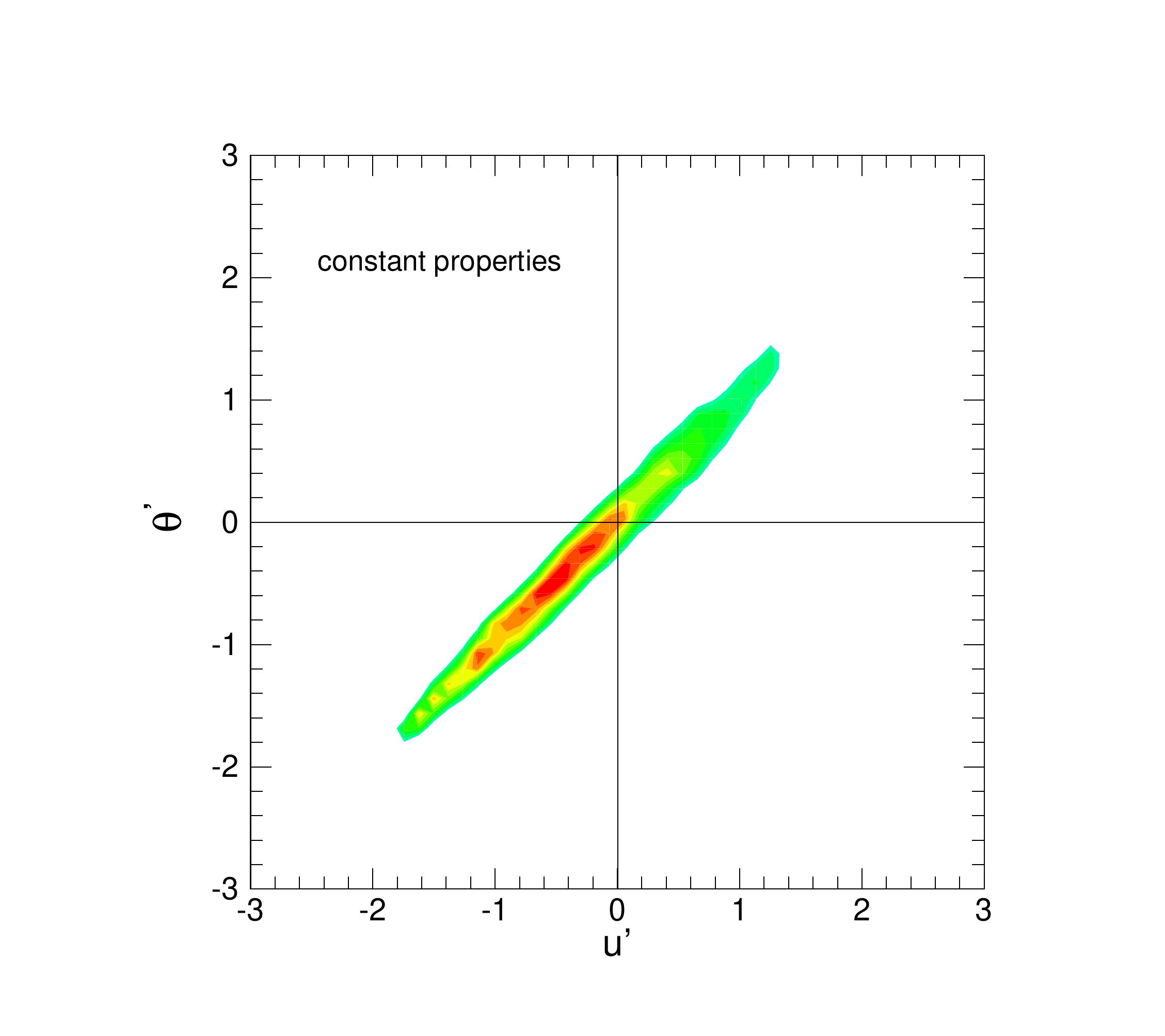}
      \label{fig: 12e}}
\subfloat[]{
      \includegraphics[width=0.48\linewidth]{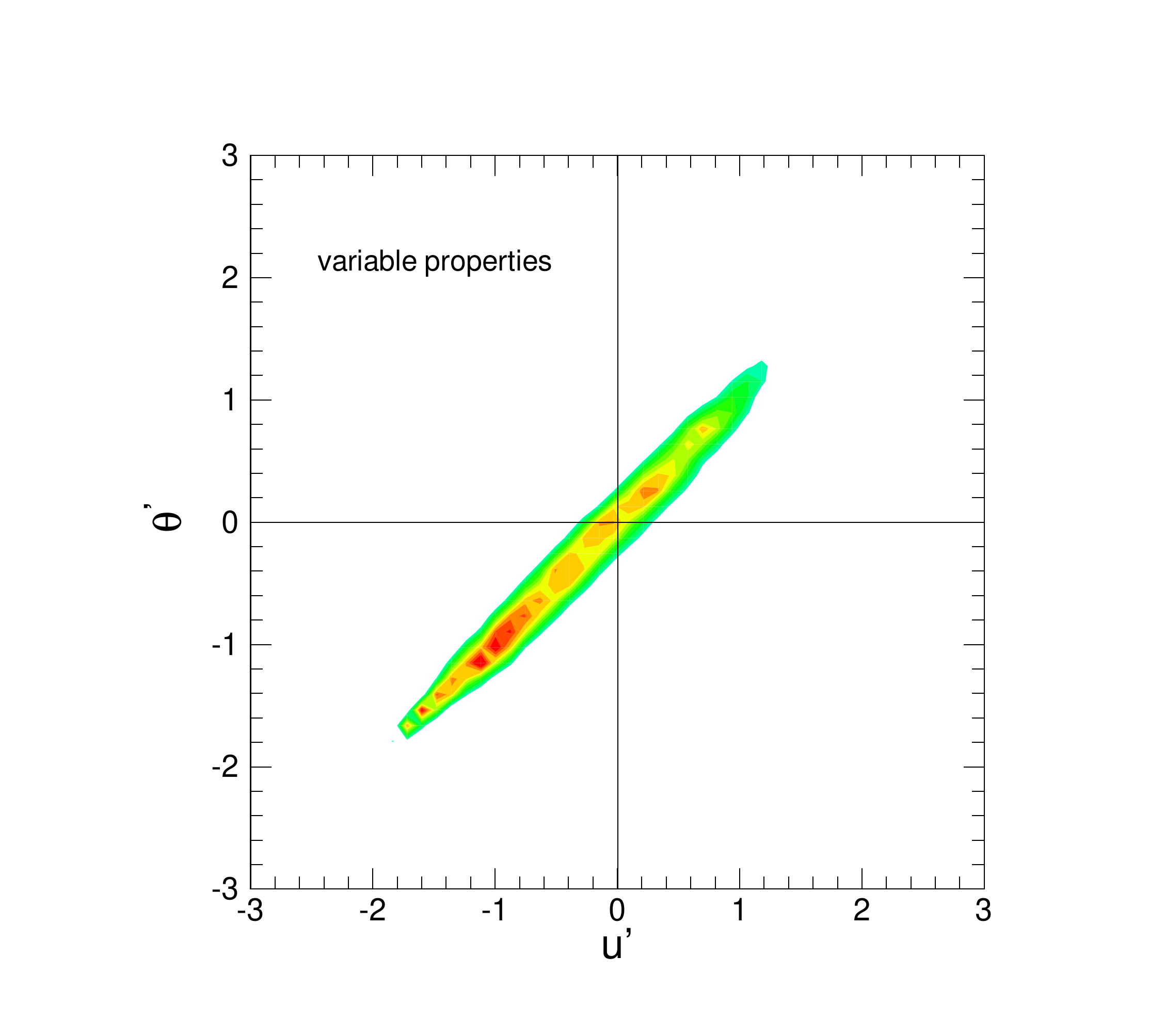}
      \label{fig: 12f}}
   \caption{Joint probability distribution function at $y^+ = 5.5$.}
   \label{fig: 12}
\end{center}   
\end{figure}
The joint PDFs (JPDFs) are very valuable in describing the physics of turbulent flows. At $y^+ = 5.5$,  the JPDFs of the streamwise and wall-normal velocity component fluctuations $u'$ and $v'$ are presented in Fig. \ref{fig: 12a}  and Fig. \ref{fig: 12b}, respectively. Here, the fluctuating values used for JPDFs are scaled with respect to the rms value. Under the constant properties, Fig. \ref{fig: 12a}  demonstrates that the second quadrant ($u' < 0$ and $v' > 0$, ejections) and fourth quadrant ($u' >0$ and $v' < 0$, sweeps) have larger contributions to the Reynolds shear stress and thus to turbulent energy production. This is in agreement with most existing studies. For variable properties, Fig. \ref{fig: 12b} shows that the ejection events are slightly less vigorous. Similarly, we observe that ejections ($v' > 0$, $\theta' < 0$) and sweeps ($v' < 0$, $\theta' > 0$) are dominant motions in contributing to the wall-normal heat flux, as shown in Fig. \ref{fig: 12c}. This corroborates the DNS results by Kong \emph{et  al.} \cite{kong2000}. Under the influence of variable properties, as shown in Fig. \ref{fig: 12d},  this trend is almost unaffected. As we discussed earlier, the similarity between the Reynolds shear stress $<-u'v'>$ and wall-normal heat flux $<-v'\theta'>$ indicates the strong coupling between streamwise velocity $u'$ and temperature $\theta'$. Fig. \ref{fig: 12e} does  demonstrate such a strong correlation and further supports this conclusion. Under the influence of variable properties, Fig. \ref{fig: 12f} illustrates that the peak values of JPDF are somewhat attenuated. Nevertheless, the magnitude of $<u'\theta'>$ at this position is found  to be  unaffected, as shown in Fig. \ref{fig: 7c}.
\subsection{Distribution of velocity and temperature increments}
\begin{figure}[!h]
\begin{center}
\subfloat[]{
      \includegraphics[width=0.5\linewidth]{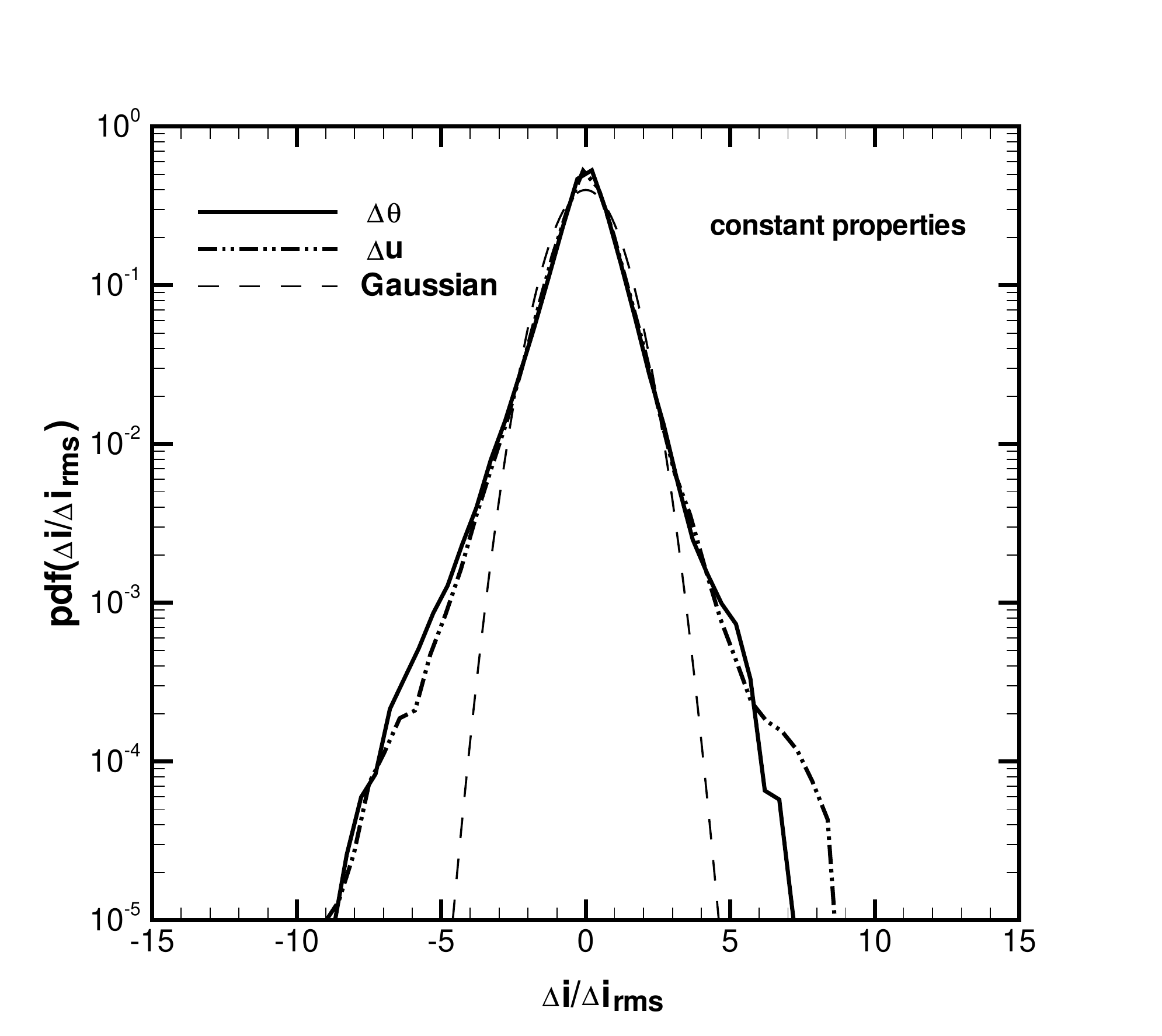}
      \label{fig: 10a}}
\subfloat[]{
      \includegraphics[width=0.5\linewidth]{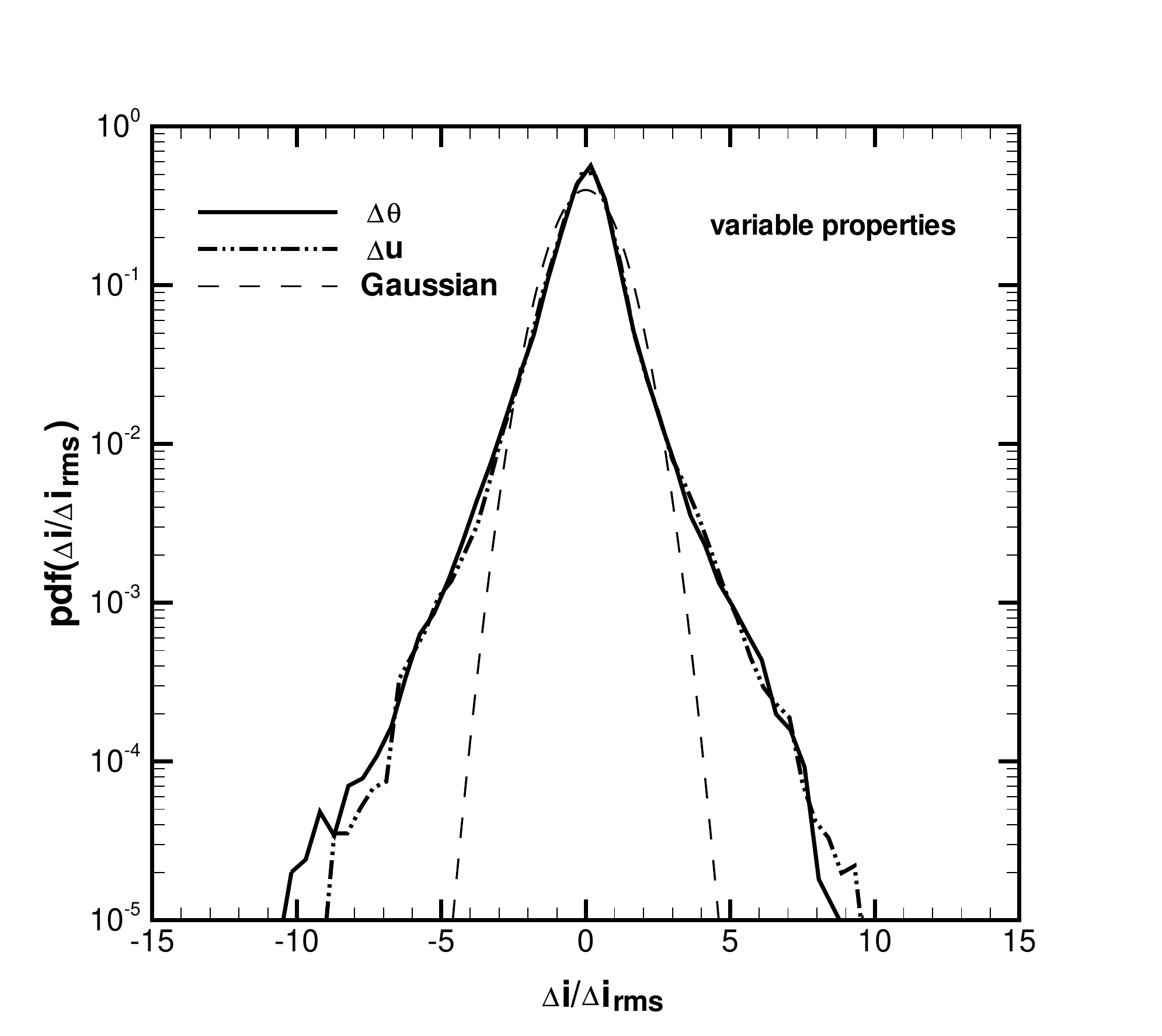}
      \label{fig: 10b}}
      
\subfloat[]{
      \includegraphics[width=0.5\linewidth]{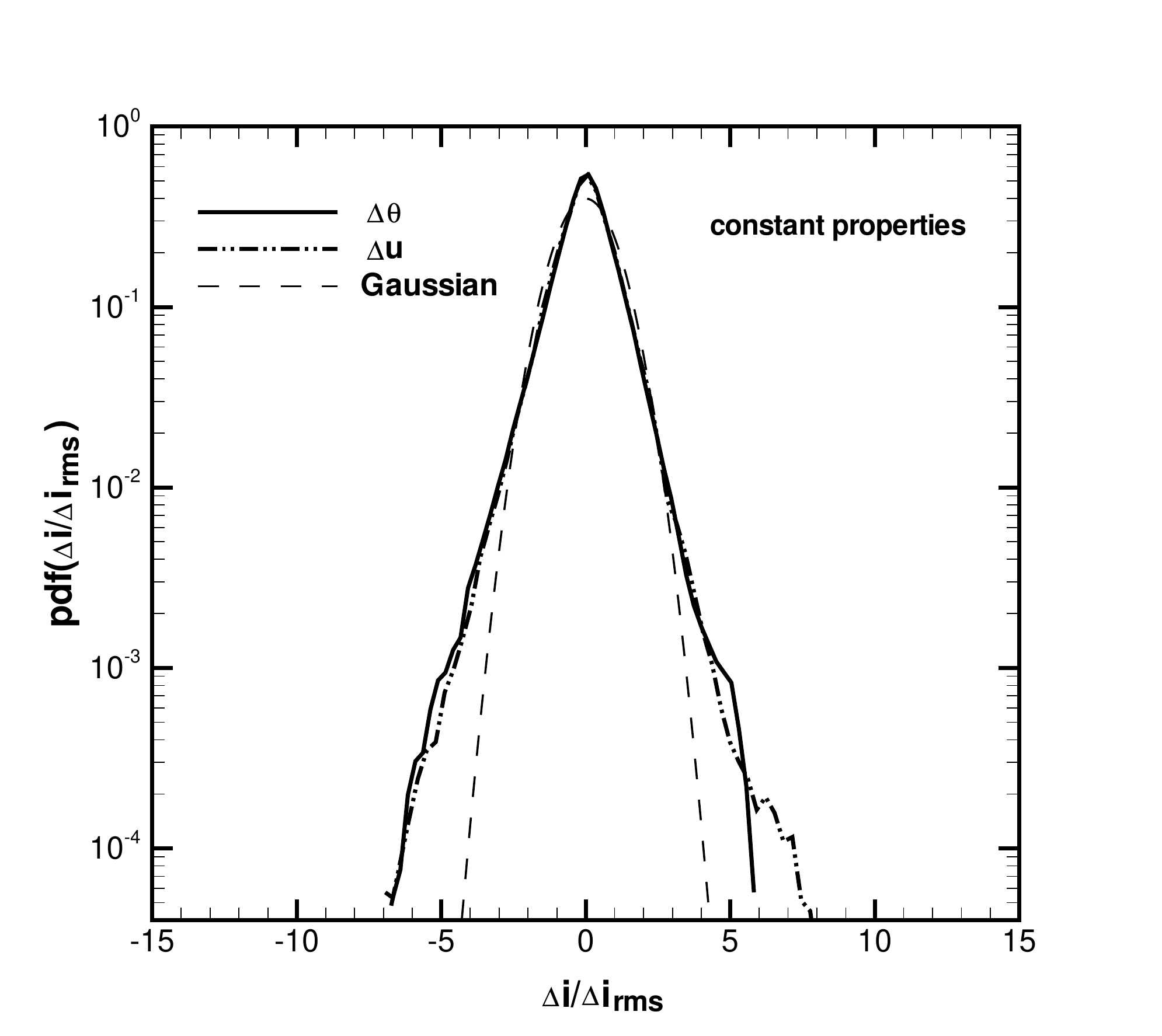}
      \label{fig: 10c}}
\subfloat[]{
      \includegraphics[width=0.5\linewidth]{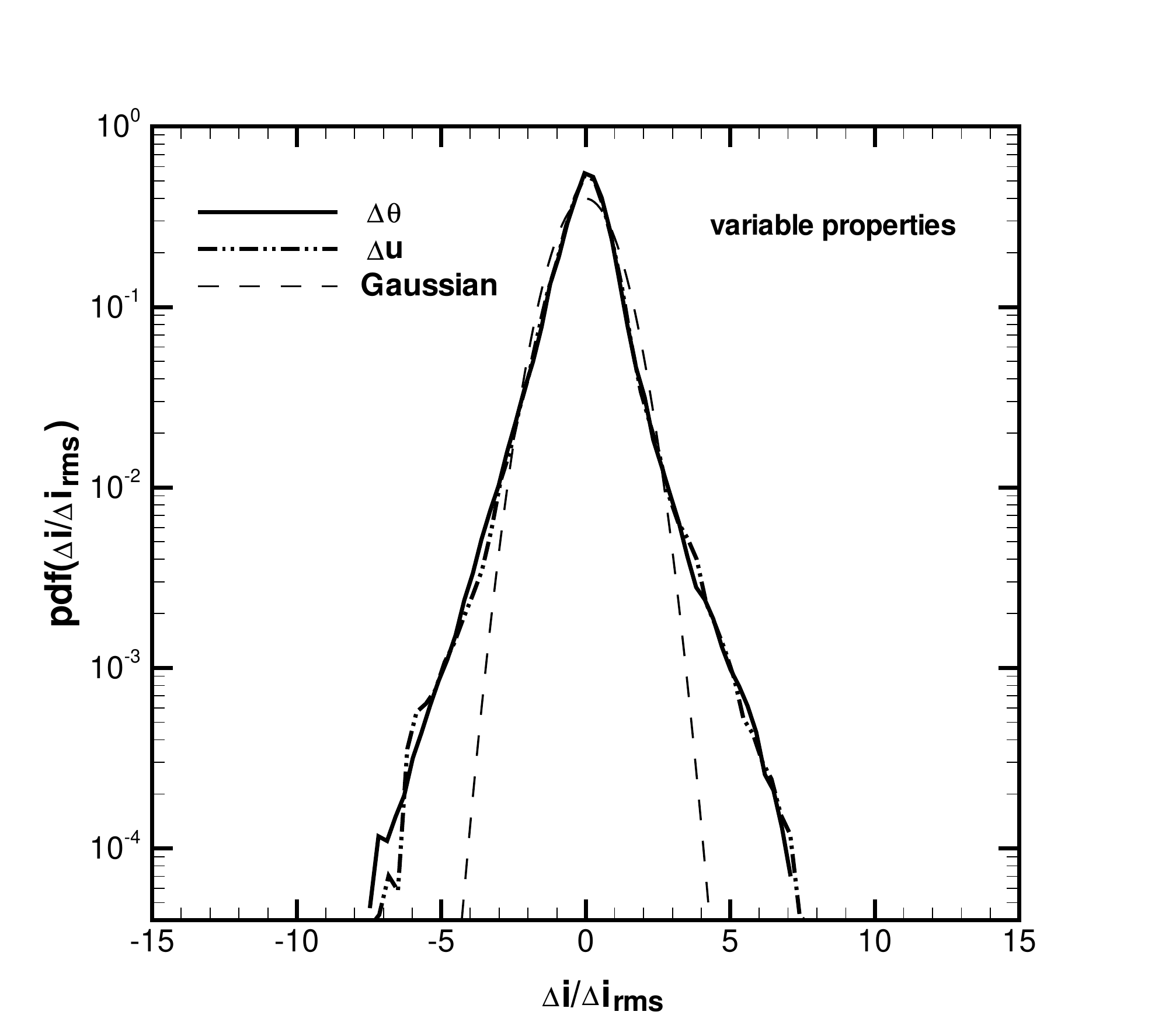}
      \label{fig: 10d}}
   \caption{Probability distribution of second order structure function for (a), (b) $r \approx 3\eta$; (c), (d) $r= 15\eta$. Here $r$ is the distance between two points and $\eta$ is the Kolmogorov scale. }
   \label{fig: 10}
\end{center}   
\end{figure}

In order to further verify the strong coupling between streamwise velocity and temperature, we calculate the PDFs of the increments of $u$ and $\theta$, respectively. Here, the increments are denoted as $\Delta u =  u(x+r)-u(x)$, and  $\Delta \theta= \theta(x+r)-\theta(x)$, where $r$ represents the  distance between two given points. Usually, the separation $r$ is located in the inertial sub-scale, \emph{i.e.}, $\eta < r < \delta$, where $\delta$ is the half-height of the channel, and $\eta$ is the Kolmogorov scale which is estimated by the relation $\eta = \delta/Re_\tau^{3/4}$. 
The PDFs of both increments at the position $y^+ = 5.5$ are plotted in Fig. \ref{fig: 10}, where the separation $r=3\eta$ and $15\eta$, respectively. It is clear for both constant and variable properties, the distribution of $\Delta u$ and $\Delta \theta$ are in good agreement. For smaller separation at $r=3\eta$, however, the PDF profiles are characterized by wider tails. This indicates the small-scale intermittency which has been observed in various turbulent flows \cite{sreenivasan1997,onorato2000}. A proper understanding of small-scale turbulence is especially important for practical calculations such as LES and RANS models. At a larger scale  $r=15\eta$, it is noted that the wider tails become narrower, and the curves approach the Gaussian distribution. If $r = \delta$, for example, we noticed that the PDFs of increments are essentially indistinguishable from a Gaussian (not shown here).
\subsection{Budget terms}
Concerning thermal flows, the equation for the budgets of the turbulent heat fluxes, $\left< u_i \theta\right>$, and temperature variance, $ k_\theta = \left <\theta^2 \right >$, are referred to Kawamura \textit{et al.} \cite{kawamura1999}. These equations are transcribed here to facilitate the reading of the paper.

The budget equation for the turbulent heat fluxes, $\left< u_i \theta\right>$, is given by:
\begin{equation}
     \frac{D \left< {u'_i}^+{\theta'}^+\right>}{D t^+} = \underbrace{P_{i\theta}}_{\text{\tiny{Production}}} -\underbrace{\epsilon_{i\theta}}_{\text{\tiny{Dissipation}}} + \underbrace{T_{i\theta}}_{\text{\tiny{Turbulent diffusion}}} +\underbrace{D_{i\theta}}_{\text{\tiny{Molecular diffusion}}}+\underbrace{\Pi_{i\theta}}_{\text{\tiny{Pressure-temperature gradient correlation}}}
    \label{Eq: 8}   
\end{equation}
where $D/Dt$ is the material derivative. The different terms on the right-hand side respectively are defined as follows:
\begin{equation}
 P_{i \theta} = -\left<{u_k'}^+{\theta'}^+\right> \frac{\partial \left<{U_i}^+\right>}{\partial x_k^+}-\left<{u_i'}^+{u_k'}^+\right>\frac{\partial\left<\Theta^+\right>}{\partial x_k^+}+\left<{u_1'}^+{u_i'}^+\right>\frac{\partial \left<T_w\right>}{\partial x_1^+} 
     \label{Eq: 9}  
\end{equation}
\begin{equation}
\epsilon_{i\theta} = \left( 1+\frac{1}{Pr}\right)\left<\frac{\partial \theta'^+}{\partial x_k^+}\frac{\partial u_i'^+}{\partial x_k^+}\right>    
    \label{Eq: 10}  
\end{equation}
\begin{equation}
T_{i\theta} = -\frac{\partial}{\partial x_k^+}\left< {u'_i}^+ {u'_k}^+{\theta '}^+ \right>
    \label{Eq: 11}
\end{equation} 
\begin{equation}
\Pi_{i\theta} = -\left<\theta '\frac{\partial {p'}^+}{\partial x_i^+} \right>
    \label{Eq: 12}  
\end{equation}
\begin{equation}
D_{i\theta} = \frac{\partial}{\partial x_k^+}\left(\left< {\theta '}^+\frac{\partial {u'_i}^+}{\partial x_k^+}\right> +\frac{1}{Pr}\left< {u'_i}^+\frac{\partial {\theta '}^+}{\partial x_k^+}\right>\right).
    \label{Eq: 13}  
\end{equation}
The budget terms of the temperature variance are expressed as follows:
\begin{equation}
    \frac{D k_\theta^+}{D t^+} = \underbrace{P_\theta}_{\text{\tiny{Production}}} + \underbrace{T_\theta}_{\text{\tiny{Turbulent diffusion}}} + \underbrace{D_\theta}_{\text{\tiny{Molecular diffusion}}} -\underbrace{\epsilon_\theta}_{\text{\tiny{Dissipation}}}
    \label{Eq: 14}
\end{equation}
The different terms on the right-hand side are defined as follows:
\begin{equation}
    P_\theta = -\left<{u_k'}^+{\theta'}^+\right> \frac{\partial \left<{\Theta}^+\right>}{\partial x_k^+}+\left<{u_1'}^+{\theta '}^+\right>\frac{\partial \left<T_w\right>}{\partial x_1^+}  
    \label{Eq: 15}
\end{equation}
\begin{equation}
T_{\theta} = -\frac{1}{2}\frac{\partial}{\partial x_k^+}\left< {u'_k}^+{{\theta '}^+}^2 \right>
    \label{Eq: 16}
\end{equation}
\begin{equation}
D_{\theta} = \frac{1}{2Pr}\left<\frac{\partial ^2{\theta '^+}^2}{\partial {x_k^+}^2}\right>.
    \label{Eq: 17}  
\end{equation}
\begin{equation}
\epsilon_{\theta} = \frac{1}{Pr}\left<\frac{\partial \theta'^+}{\partial x_k^+}\right>^2    
    \label{Eq: 18}  
\end{equation}
 \begin{figure}[!h]
\begin{center}
\subfloat[]{
\includegraphics[width=0.65\linewidth]{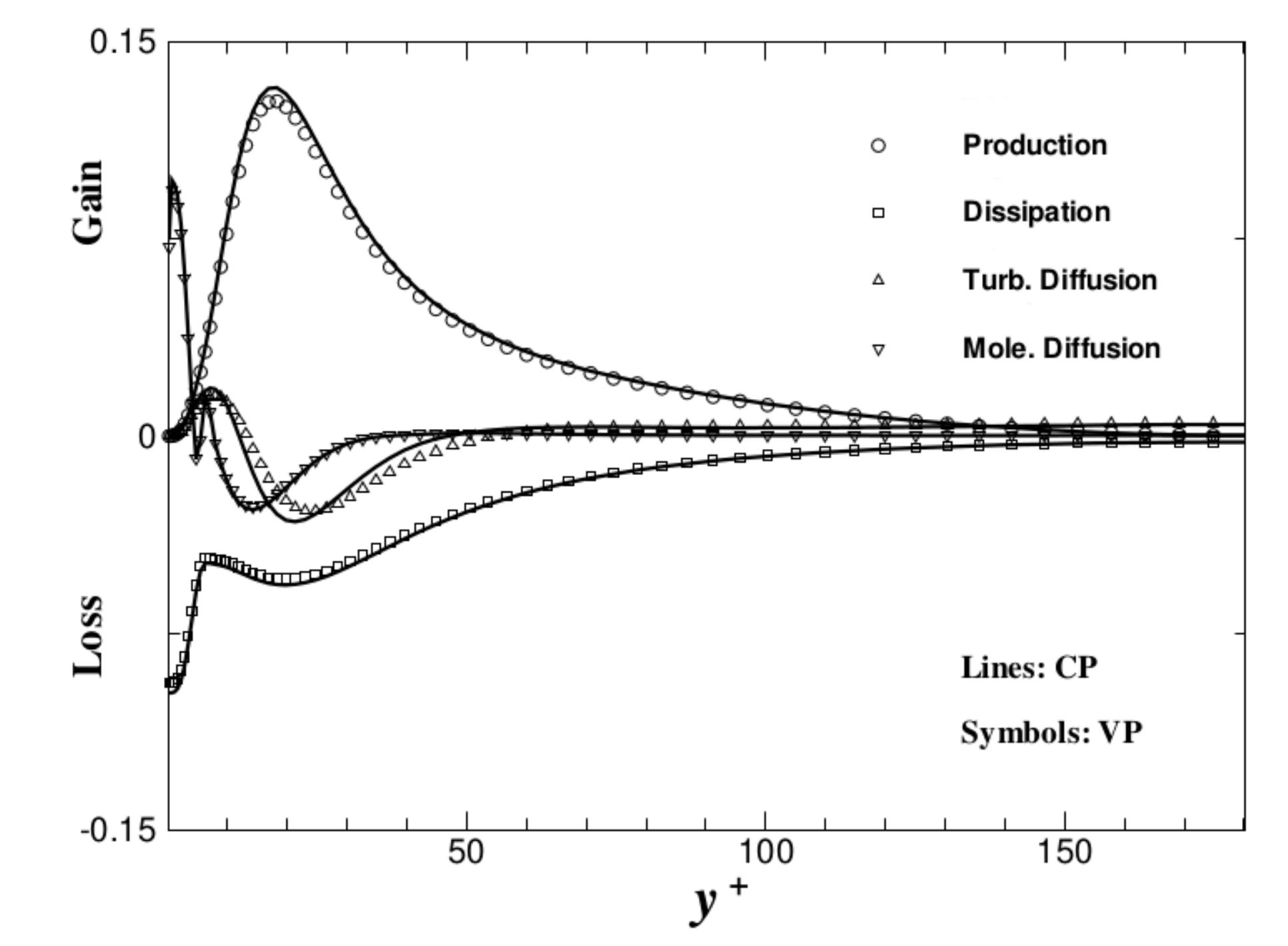}
      \label{fig: 13a}}
      
\subfloat[]{
\includegraphics[width=0.65\linewidth]{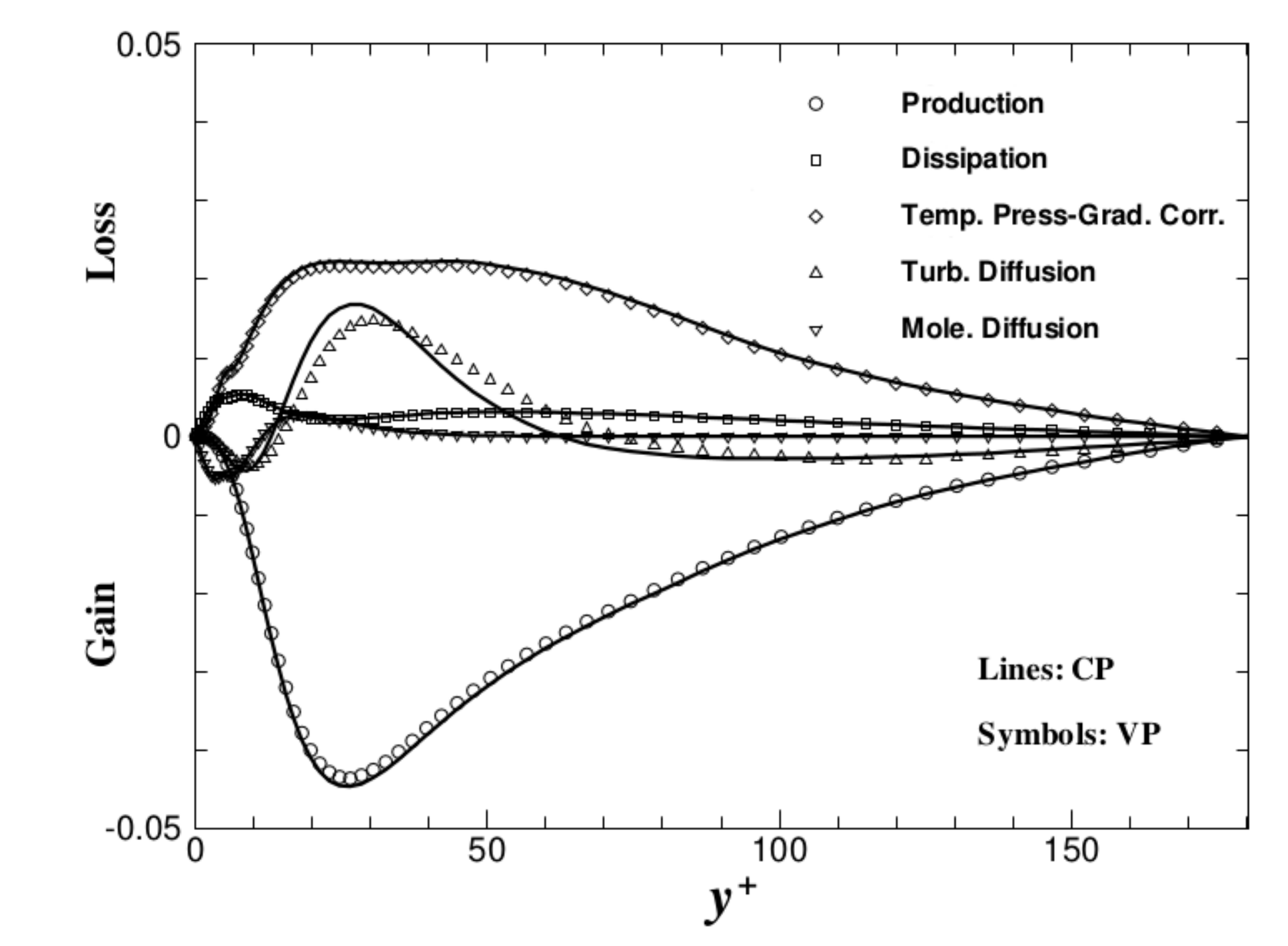}
      \label{fig: 13b}}
   \caption{Budget terms for (a) temperature variance and (b) wall-normal heat flux.}
   \label{fig: 13}
   \end{center}
\end{figure} 
 The effect of the variable properties in the budget terms of the temperature variance and the wall-normal heat flux is given in Fig. \ref{fig: 13} and Fig. \ref{fig: 13b}, respectively. Compared to the DNS study of Kawamura \textit{et al.} \cite{kawamura1999}, the essential features of various budget terms are captured. In particular, for the production term of temperature variance and wall-normal heat flux, the peak position is appropriately predicted while the peak value is underestimated. In terms of the influence of the variable properties on the budgets, Fig. \ref{fig: 13a} and Fig. \ref{fig: 13b} show that only the turbulent diffusion term is slightly modified, while the other terms almost remain unaffected, as  different budgets for variable properties collapse well with the benchmark case. 
 \newpage
\section{Conclusions}
\label{sec:4}
In this paper, the influence of temperature-dependent viscosity and thermal conductivity on the forced convection in a channel is investigated by the wall-resolved LES. Different from the isothermal boundary condition, the present isoflux boundary condition permits the wall temperature fluctuations, thus induces the variations of fluid property. To highlight the influence of variable properties on turbulent flow, which is inversely proportional to the Reynolds number, we select $Re_\tau = 180$ in this study. Although the variable viscosity results in a lower bulk Reynolds number than the benchmark case with constant properties, the two cases are considered to be fully turbulent flow. 

Using the classic wall scaling, we observed that the variable properties have a trivial effect on the mean velocity and temperature profiles. Meanwhile, the Reynolds shear stress and wall-normal heat flux, which play a fundamental role in transporting momentum and heat in wall-bounded turbulence, remain almost unaffected by the variations of property. However, we also found that the turbulence intensities, which are represented by the rms values of velocity fluctuations and temperature, are modulated in a perceptible way. Consistent with \cite{kong2000}, we also noticed the strong correlation between streamwise velocity $u$ and temperature $\theta$, which results in the remarkable similarity between $\left <-u'v'\right>$ and $\left <-v'\theta'\right>$ as well as the similarity between $u_{rms}^+$ and $\theta_{rms}^+$. Our study shows that this conclusion can be extended to the flow with variable properties. The PDF profiles of $u$ and $\theta$ show a good agreement, however, further inspection of the figure reveals that the negatively skewed peak is replaced by a plateau, indicating that the structures of low-speed streaks are modified by the variable properties. In addition, the effect of variable properties on the burst motions, such as ejections and sweeps, is illustrated. In terms of small-scale turbulence and intermittency, we cannot detect any obvious influence of variable properties. Finally, we calculated the budgets for the temperature variance and wall-normal heat flux, it is seen that only the turbulent diffusion term is slightly affected by the variable properties, whereas other terms remain unaffected.     

\section*{Acknowledgement}
The authors greatly appreciate the financial support provided by Vinnova under project number "2020-04 529". Computer time was provided by the Swedish National Infrastructure for Computing (SNIC), partially funded by the Swedish Research Council through grant agreement no. 2018-05973.







\end{document}